# Lunar Secondary Craters and Estimated Ejecta Block Sizes Reveal a Scale-dependent Fragmentation Trend

**Kelsi N. Singer[1], Bradley L. Jolliff[2], and William B. McKinnon[2]**

[1]Southwest Research Institute, 1050 Walnut St. Suite 300, Boulder, CO 80302,USA. [2]Department of Earth and Planetary Sciences and McDonnell Center for the Space Sciences, Washington University in St. Louis, Missouri 63130, USA.

**Key Points**

1. We define the maximum secondary crater size at a given distance from a primary and estimate ejecta fragment sizes and velocities.
2. We find a steep scale-dependent trend in ejecta fragment size-velocity distributions.
3. Maximum ejecta fragment sizes fall off more steeply with increasing ejection velocity for larger primary impacts.

**Abstract**

Planetary impact events eject large volumes of surface material. Crater excavation processes are difficult to study, and in particular the details of individual ejecta fragments are not well understood. A related, enduring issue in planetary mapping is whether a given crater resulted from a primary impact (asteroid or comet) or instead is a secondary crater created by an ejecta fragment. With mapping and statistical analyses of six lunar secondary crater fields (including Orientale, Copernicus, and Kepler) we provide three new constraints on these issues: 1) estimation of the maximum secondary crater size as a function of distance from a primary crater on the Moon, 2) estimation of the size and velocity of ejecta fragments that formed these secondaries, and 3) estimation of the fragment size ejected at escape velocity. Through this analysis, we confirmed and extended a suspected scale-dependent trend in ejecta size-velocity distributions. Maximum ejecta fragment sizes fall off much more steeply with increasing ejection velocity for larger primary impacts (compared to smaller primary impacts). Specifically, we characterize the maximum ejecta sizes for a given ejection velocity with a power law, and find the velocity exponent varies between approximately -0.3 and -3 for the range of primary craters investigated here (0.83–660 km in diameter). Data for the jovian moons Europa and Ganymede confirm similar trends for icy surfaces. This result is not predicted by analytical theories of formation of Grady-Kipp fragments or spalls during impacts, and suggests that further modeling investigations are warranted to explain this scale-dependent effect.

**Plain Language Summary**

When an impact crater forms on a planetary surface, fractured pieces of the surface material are ejected and can form their own secondary craters outside of the main, or primary, crater. Secondary craters are found in large numbers on the Moon and other bodies, and they provide evidence of the impact process. The size of secondary craters is related to the size of the ejecta fragment, and the distance of the secondary crater from the primary is related to fragment ejection velocity. It was already known that the largest fragments are generally ejected at lower velocities for a given impact event; in addition, we find that fragment size shows a stronger dependence on ejection velocity for larger craters. For smaller craters, the size of the largest fragment does not change as much with ejection velocity. This information can be used for constraining models of impact physics and fragmentation of the surface, and for estimating the size of fragments ejected at escape velocity, which may later form meteorites on other bodies. We also provide an estimate of the maximum secondary crater size as a function of distance from a given primary crater, which can help distinguish between primary and secondary craters.

## 1. Introduction

When a projectile impacts a planetary surface, the resulting shock and rarefaction waves fragment and eject material (Melosh, 1989). The highest velocity fragments ejected during the early stages of crater formation may reach escape velocity and may later end up as meteorites on other planets. Alternatively, these escaped fragments could create craters of their own on the parent body or another world, termed sesquinary craters. Other fragments ejected at somewhat lower speeds re-impact the parent body creating secondary craters. The slowest material ejected during the final stages of the crater formation forms a more continuous deposit directly outside of the final crater rim. Most previous studies addressed the bulk properties of ejecta; our study uses an empirical approach to determine the details of individual ejecta fragment sizes and ejecta maximum size-velocity distributions (MSVDs) from planetary-scale impacts in geological materials, a task that is difficult in most theoretical, modelling, and laboratory studies of ejecta.

Secondary craters are recognized on many solid surfaces imaged throughout the Solar System, including the terrestrial planets, outer Solar System moons, and large asteroids such as Vesta and Ceres (Bierhaus et al., 2018; McEwen and Bierhaus, 2006; Schmedemann et al., 2014; Schmedemann et al., 2017; and references therein). Exceptions include Earth, Jupiter's moon Io, and Saturn's moon Titan (all locations where secondary craters have been obscured by later geologic action), and small bodies with low escape velocities. Planetary scientists leverage comparative planetology to understand cratering and ejecta processes across the Solar System. Studies of secondaries on bodies such as Mars and Jupiter's moon Europa, which both exhibit terrains with very few primary craters, helped illuminate the large number of secondaries produced in a given impact and their great radial extent (Bierhaus et al., 2005; McEwen et al., 2005).

We focus on Earth's Moon, as its extensive cratering record provides a natural laboratory for understanding the cratering process; one can map secondaries around a wide range of primary craters sizes, and on varied target surfaces. The high quality Lunar Reconnaissance Orbiter Camera (LROC) dataset allows for an unprecedented characterization of secondary craters with both the low sun, 100-m-px$^{-1}$ Wide Angle Camera (WAC) global mosaic and high-resolution Narrow Angle Camera (NAC) images (0.5-to-2 m px$^{-1}$). Our results can then be compared with earlier studies (see supplement section S3), and in particular Lunar-Orbiter-based studies of secondary craters on the Moon (Vickery, 1986, 1987) and our own work on icy satellites (Singer et al., 2013).

Early researchers recognized the general importance of secondaries on the Moon (Shoemaker, 1965; Shoemaker et al., 1962), and aspects of secondary cratering and ejecta played a role in a large variety of lunar scientific contexts, from relative and absolute age dating of geologic units to terrain formation and movement of material across the lunar surface. One of the important calibrations of cratering rates to absolute ages is based on secondaries from the crater Tycho (with a diameter of 86 km), which were taken as the cause of a landslide-like downslope movement of boulders and a light mantling material off the South Massif at the Apollo 17 site. The exposure ages of boulders were then used to date the Tycho impact (Arvidson et al., 1976; Lucchitta, 1977; Wolfe, 1981).

We define a secondary crater as a depression in the ground created by a fragment ejected in a primary impact. Many secondary craters are formed by ejecta traveling below the sound speed of the target body, and as such, are not necessarily created by hypervelocity impacts (additional discussion in section 4). Secondary crater characteristics can include: generally shallow depth-to-diameter ratios (*H/D*) compared with primary craters, steep size-frequency distributions, commonly elliptical or otherwise asymmetrical planform shapes, chevron, herringbone, or arrow-like ejecta patterns, and they commonly occur in radial chains or clusters pointing back to the parent crater (see extensive summary in McEwen and Bierhaus, 2006 and references therein). There are also less obvious secondary craters that are fairly circular and that may occur individually or in a loose cluster, but otherwise show little or no morphologic character of their secondary origin. In particular, distant secondaries (those landing at higher velocity) become more circular, and older secondaries eventually lose their distinctive ejecta patterns (through subsequent micrometeorite bombardment and small crater gardening, similar to degradation of other topography or bright rays (Fassett and Thomson, 2014; Honda et al., 2012; Wilhelms et al., 1987)).

Because *H/D* varies for both small primary and secondary craters (Basilevskii, 1976; Basilevsky et al., 2014; Chappelow, 2018; Mahanti et al., 2014; Mahanti et al., 2018; Stopar et al., 2017), a shallower crater is not in itself diagnostic of a secondary crater, but may be a helpful indicator. Surface roughness or maturity measures—such as radar circular polarization ratio (Wells et al., 2010), statistics related to topographic profile curvature (Kreslavsky et al., 2013), ultraviolet reflectance ratios (Denevi et al., 2014), or even simply crater densities (Xiao et al., 2014)—can reveal the asymmetrical ejecta of very fresh secondaries or linear features associated with secondary rays and clusters.

With the current study we define an upper bound on secondary size at a given distance from a primary crater (Section 3). We use this relationship to help identify secondaries across the lunar surface, and also to test hypotheses involving assumptions about secondary craters. For the estimated ejecta fragment sizes (Section 4), we find a scale-dependent trend not previously predicted by impact fragmentation theories (Grady and Kipp, 1980; Melosh, 1984; 1989, p. 107). Implications for age dating of planetary surfaces, formation of meteorites, and modelling of cratering as a Solar System-wide geologic process are discussed (Section 5).

## 2. Image Data Sources and Mapping Methods

The LROC Wide Angle Camera 100-m-px$^{-1}$ global mosaic with incidence angles 55°–75° (suitable for mapping based on topography (Speyerer et al., 2011)) served as the base for mapping of secondary fields around primary craters above 30 km in diameter. We also examined Narrow Angle Camera (NAC) images (~0.5–1.5 m px$^{-1}$) for confirmation of secondary crater morphologies and for mapping around the three smaller primary craters (0.8–3.3 km diameter). High-Sun mosaics (available on the LROC website) were useful for following crater rays. The images were processed in the USGS ISIS program and mapped in ArcGIS (all length measurements were made geodesically). Information on image sources and mapping of the icy satellite secondary crater data, shown below for comparison with similar data for the Moon, can be found in *Singer et al.* [2013].

We map and mathematically characterize the maximum size-range distribution (MSRD)

of secondary craters around 6 primary craters on the Moon (Table 1; Section 3) ranging between 0.8-to-660 km in diameter ($D$). The primary craters mapped here are relatively young and well-preserved for their size, and thus many secondaries are easily recognizable. During the mapping phase, all craters were coded with a confidence level based on how many expected morphologic characteristics of secondary craters each crater displayed. The morphologies of well-defined, or so called "obvious" secondary craters can include: v-shaped or chevron-like ejecta, elongation in the radial direction, asymmetrical rim heights (most often with a less well-defined rim in the downrange direction), and occurrence in a chain, cluster, or ray of craters that share these morphologies. Radial great circles extending from each primary crater were used to check the alignment of rays, clusters, and chains of secondary craters, as well as the v-shaped ejecta or elongation of the secondary craters themselves. The more distant secondaries are found along rays that can be traced from the primary crater. Additionally, the degradation state (or freshness) of secondaries in a given field is generally similar across the field, and this aspect was used where possible to distinguish secondaries from earlier or later cratering events (with the caveat that lighting geometries must be fairly consistent in order to assess this). We discuss secondary crater depths in section 4.1, but depth-to-diameter ratios were not used as a selection criterion in this study. Each secondary field was re-examined several times throughout the mapping phase to promote consistency of mapping and identification across the project.

The goal was not to map every possible secondary crater down to the resolution limit, as we were primarily focusing on *characterizing the maximum secondary size* at a given distance. Additionally, our study goals were derived to fit with the reality of mapping secondary craters: it is impossible to map all secondary craters due in part to their clustered/overlapping nature, thus we do not focus on achieving completeness as is often done for studies of primary crater populations. For all secondary fields, only craters with the highest likelihood of a secondary origin (those in obvious radial chains or clusters, and with at least two of the morphological characteristics described above) and with well-defined rim walls were used in the subsequent analysis.

**Table 1. *Primary Crater and Secondary Field Characteristics***

| Primary Crater | Primary diameter (km)[a] | Primary transient diameter (km)[b] | Primary impactor diameter (km)[b] | Number of secondaries used in the analysis | Largest observed secondary (km)[c] | Average of largest 5 secondaries (km)[c] | Estimated maximum fragment size at escape velocity (m)[d] |
|---|---|---|---|---|---|---|---|
| Orientale | 660 | 360 | 85 | 245 | 26 (4%) | 23 (4%) | 860 |
| Copernicus | 93 | 63 | 9.3 | 4,565 | 5.5 (6%) | 4.9 (5%) | 50 |
| Kepler | 31 | 24 | 2.7 | 1,205 | 1.4 (5%) | 1.3 (4%) | 40 |
| Unnamed in SPA[e] | 3.0 | 2.5 | 0.16 | 1,884 | 0.18 (5%) | 0.16 (5%) | 3 |
| Unnamed near Orientale[e] | 2.2 | 1.8 | 0.11 | 2,645 | 0.10 (5%) | 0.08 (4%) | 5 |
| Unnamed in Procellarum[e] | 0.83 | 0.68 | 0.038 | 1,728 | 0.04 (5%) | 0.04 (5%) | 5 |

[a]Final diameter for Orientale is estimated at the Outer Rook Mountains.

[b]Calculations described in section 4.6.

[c]Percentage of the primary diameter given in parentheses and examples of largest secondaries are shown in Figs. 1, S3-7.

[d]Fragment sizes are estimated with quantile regression fit parameters as in Fig. 9c,d and Table S2. See additional information in Section 4.

[e]See Figs. 5–7 for additional information about the smaller, unnamed craters.

Combined for all of the six secondary fields, 10s of thousands of features were considered, but only the highest confidence category was retained for analysis, yielding 12,385 high confidence secondary craters (see Tables 1 and S4). We focused on capturing as many of the largest and most distinct secondary craters (at a given distance) as possible, while still capturing many of the smaller ones down to the image resolution limits. This means that in in any given cluster or chain, typically more of the large features are captured, and only a fraction of the smaller features are mapped because there are many overlapping features that cannot be easily distinguished. In the transition zone between the proximal continuous ejecta deposits (sometimes referred to as the ejecta blanket) and the distinct secondary crater field, secondary craters were mapped when they appeared to be an approximately ellipsoidal depression that could plausibly be from a distinct fragment, rather than a less distinct radial groove or low topographic amplitude feature with a general herringbone pattern. It is possible that a pre-existing crater can be scoured by an ejecta deposit, giving it an elongated appearance. But this process is not, to our knowledge, thought to form v-shaped ejecta around an individual crater. We cannot rule out that a few primary craters were incorrectly marked as secondaries, but the criteria utilized here were intended to focus the results on only those craters with the highest likelihood of being secondary craters.

Secondary crater diameters were generally measured perpendicular to the direction radial to the primary crater, as the downrange secondary crater walls are often less distinct. Where the WAC basemap was used as the main mapping dataset, we also reviewed the available NAC images for each secondary field to confirm the smaller morphological aspects such as v-shaped ejecta.

The secondary craters from Orientale are somewhat different from the other fields because they are so large in size, and are relatively old. In this case, it is impossible to map secondary craters down to the resolution limit, so again we focus on the largest secondaries. Hints of v-shaped ejecta and often the asymmetric wall heights are still apparent for many craters in chains and clusters around Orientale. However, the indicators are not as strong in the Orientale field and it is less easy to distinguish between secondaries and primaries.

We focus on secondaries that are more likely to have been formed by a single ejecta fragment (or tight cluster, see below) because we use the size of the secondary crater and scaling laws to estimate the mass and size of the fragment that formed each secondary (Section 4). It is possible for a packet of tightly clumped fragments to produce a final crater similar in morphology to one made by a more coherent "single" fragment, but the crater morphology does become more distinctive as the fragment dispersion increases (Schultz and Gault, 1985; Shuvalov and Artemieva, 2015). Specifically, the increase in dispersion of a cluster of fragments leads to shallower craters with more irregular topography and sometimes multiple distinct herringbone-style streaks. Schultz and Gault (1985) point out some observed secondary craters have features similar to these and may be formed by clustered impactors. Distinct clusters of boulders have been observed in the continuous ejecta deposits of the terrestrial Lonar Crater (Kumar et al., 2014). Although we cannot determine uniquely what occurred for each of the mapped secondaries, we attempted to only map features that were less distended and irregular. The features used for analysis here do not look like those produced in the more dispersed examples of Schultz and Gault (1985). If some of the secondaries mapped here were formed by a tight cluster of fragments (such that we cannot distinguish them morphologically) these fragments could have been ejected as a more coherent block and broken up in flight, or may have been ejected as a tight cluster. Either way, they form a mass of material that was most likely ejected from the same location within the transient crater, at the same time, and with the same velocity (in order for them to form a single, distinct secondary crater).

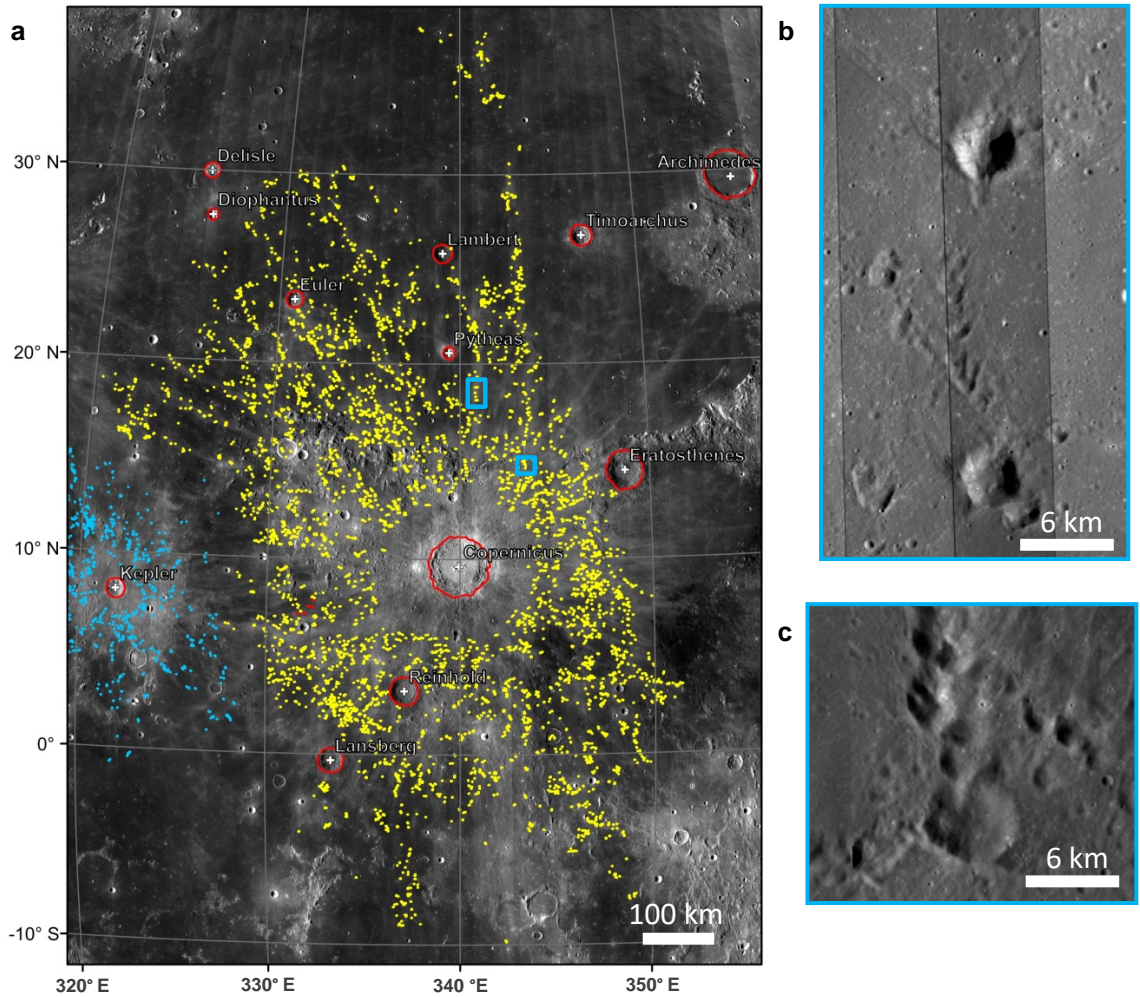

**Figure 1. Secondary crater fields. a,** Secondary crater distribution around Copernicus (93 km in diameter; yellow points) and Kepler (31 km; blue points). Shown on WAC global mosaic, 100 m px⁻¹ (Speyerer et al., 2011), with orthographic projection centered on Copernicus (9.5°N, 340.0°E). Only clearly defined secondaries are marked. Secondaries are most easily identifiable in the prominent rays. The red points west of Copernicus indicate volcanic craters, and blue outlines show the location of large secondaries shown in b and c. **b and c,** Detailed views of secondary craters. NAC pairs M1108403627L,R and M1139030009L,R (1.3 m px⁻¹).

## 3. Secondary size-range relationship

The mapped secondary crater field for Copernicus is shown in Fig. 1, and secondary sizes are plotted against their distance from the 93-km-diameter primary crater in Fig. 2. These data exhibit a well-known trend: secondary crater diameters ($d_{sec}$) generally decrease with increasing distance from the primary, related to the ejection of smaller fragments at higher velocity during primary crater excavation (Melosh, 1984, 1989; Vickery, 1986; Vickery, 1987). We use Copernicus as our example field to explain the methodology and explore the scaling parameter space (described below) because the dataset includes the most measurements ($n = 4{,}565$) and it is possible to follow Copernicus' rays for great distances (secondaries are measured up to 870 km distance). Full maps for the other five secondary fields are presented in Figs. 3-7 and results for all secondary fields are shown in Figs. 8, 9, and 11.

At any given range ($R$) from the primary, secondary craters occur at a maximum size ($d_{sec,max}$) and at many smaller sizes, down to the resolution limit of the images. Even the smallest primary crater surveyed here, an extremely fresh, unnamed 0.8-km-diameter primary crater in Oceanus Procellarum (Fig. 7), exhibits very small proximal secondary craters, down to at least 10 m in diameter, and many smaller craters near the resolution limit (~5 m in diameter) also appear to be secondary features. We note that many unresolved, dark or light splotches occur in association with even smaller, new craters formed since the Lunar Reconnaissance Orbiter (LRO) achieved lunar orbit (e.g., splotches around an 18-m-diameter crater formed on 17 March 2013 (Robinson et al., 2015; Speyerer et al., 2016)). These bright/dark splotches are likely the result of relatively small ejecta fragments impacting and ejecting some material themselves, although image resolution is insufficient to discern if they create well-defined secondary craters or compression features, or simple debris spatters.

More discussion on these smallest craters can be found in section 4.

We characterize the upper envelope of the distribution as a constraint on the maximum secondary crater size at a given distance/range from a primary, assuming all smaller sizes are possible at the resolution of our images. Quantile regression (Koenker, 2005) fits to a power law function ($d_{sec,max} = aR^{-b}$) of the 99th quantile are given in Table 2, where $R$ is in km. Quantile regression analysis was carried out in the program STATA on the log of the data. Quantile regression is a type of regression, but instead of fitting the mean by minimizing the sum of squared residuals of each point as some linear regressions do, it minimizes the sum of a different function of the residuals (a tilted absolute value function), which allows the median or other quantiles to be fit. Bootstrap standard errors are reported (1000 repetitions). The bootstrap follows a standard procedure, where it generates 1000 datasets of the same size as the original, by sampling the data with replacement. We then calculate the standard deviation ($\sigma$) of the power law parameters for the bootstrap datasets, which represents a measure of the standard error for these parameters (Table 2). The slope parameter ($-b$ or $-\beta$) and its error do not change in the conversion from the log values to actual values. The constant calculated with quantile regression in log space [$\ln(a)$ or $\ln(\alpha)$] is converted to linear space ($a$ or $\alpha$; see Table 2) but the error term cannot be translated in a meaningful way with standard methods (e.g., the delta method) because the constant calculated in log values represents the value of $\ln(y)$ when $\ln(x)=0$, and the logarithm function is not defined at $x=0$. Additionally, the parameter values from the bootstrap draws are no longer normally distributed when converted to the actual values. Thus, we additionally calculate the confidence intervals around the 99th quantile fits as a measure of the uncertainty in the supplement section S4.

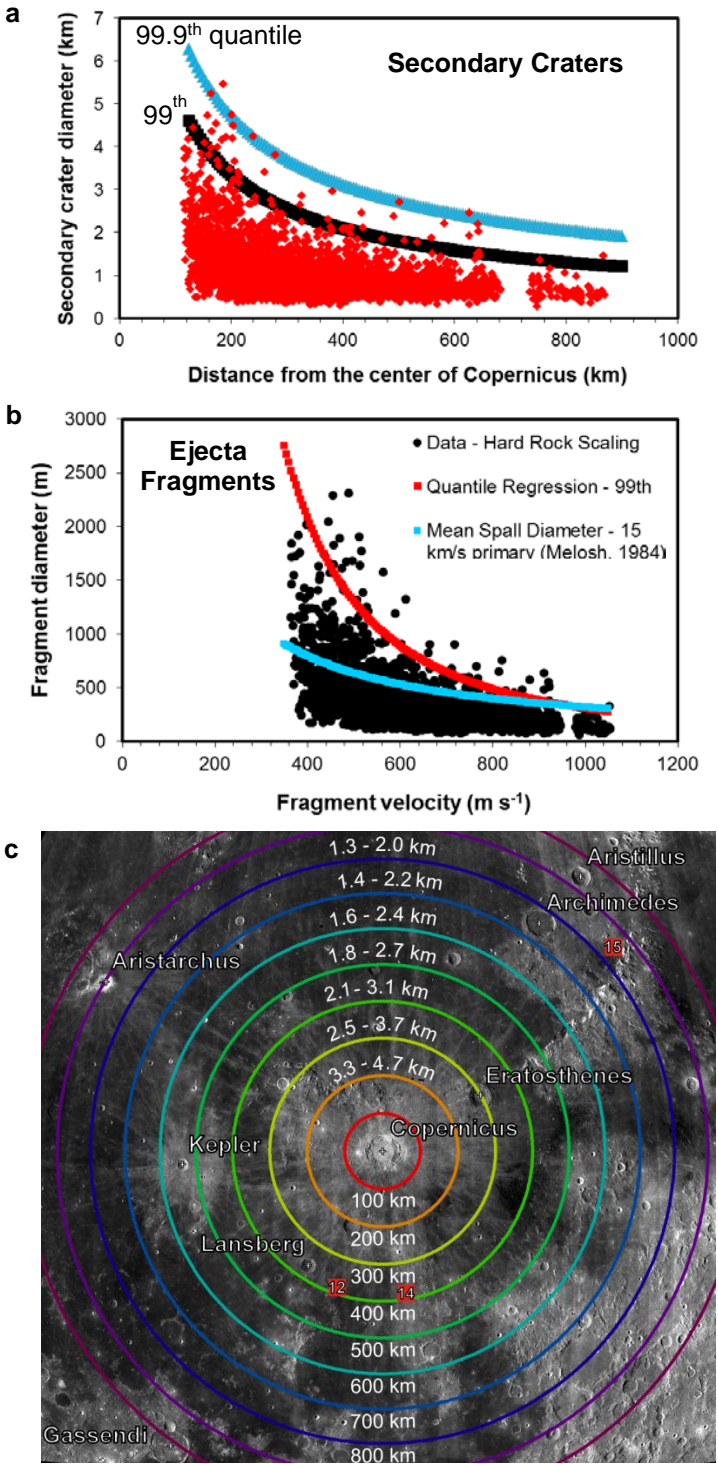

**Figure 2. Secondaries and estimated fragments for Copernicus. a,** Secondary crater diameters as a function of distance; quantile regression fits to the upper envelope of the distribution. **b,** Estimated fragment sizes (hard rock target material), quantile regression fit to upper envelope, and estimated mean spall diameter for a primary projectile impacting the Moon at 15 km s$^{-1}$ [Melosh, 1984, 1989]. See calculation details in Section 4. **c,** Maximum size of a secondary crater at 100 m radial interval ranges from Copernicus, as derived from quantile regression fits to the 99[th] and 99.9[th] quantile (parameters given in Table 2). Contours shown over LROC WAC 100 m px$^{-1}$ basemap and red squares indicate Apollo sites.

We focus here on obtaining a function that represents the upper envelope of the data. The 99[th] quantile represents a typical large secondary size, while the 99.9[th] is closer to an absolute maximum for the Copernicus secondary crater field (Fig. 2a). We give the results for several quantiles for Copernicus as an example, because quantile regression is sensitive to the bulk of the data, and each secondary field mapped here has its own characteristics. These curves estimate maximum secondary sizes at any radial distance from Copernicus; for example, at the Apollo 12 and 14 sites, Copernicus' secondaries could be as large as 2-3 km (Fig. 2c).

Power-laws were fit to the upper envelope of the data for all six secondary fields using the method noted above (Table 2). The regression parameters (*a* and *–b*) for all six fits are plotted in Fig. 9 as a function of the primary crater size. As expected, the size of the secondary craters, described by the magnitude parameter *a* (Fig. 9a), increases with increasing primary crater size (Melosh, 1989; Vickery, 1986; Vickery, 1987). A trend not predicted by impact fragmentation theories (discussed more in section 5 in reference to the ejecta fragments) emerges for the range exponents (*–b*; Fig. 9b). Secondary crater fields around large primaries have steeper falloffs than those for smaller primaries, meaning that the sizes of secondaries fall off more quickly with distance than for smaller primaries. Fit trend lines to derive equations for −b and a as a function of D and R are given in Table 3. These results are least constrained above D = 100 km. No primary craters with D >150 km but smaller than Orientale (D ~660 km) were found that exhibit well-preserved secondary fields suitable for mapping. Limited data for craters on the icy Jovian satellites Europa and Ganymede show a similar trend in −b though, and the MSRD of the ~550 km impact basin Gilgamesh (Ganymede's largest) demonstrates a statistically similar −b (Singer et al., 2013) to Orientale's. Secondary crater magnitudes (a) of icy satellites follow a similar trend with primary size.

**Table 2.** *Quantile regression fits of secondary crater size-range distributions\*.*

| Primary Crater (diameter in km) | Quantile Regression Parameters | | |
|---|---|---|---|
| *99th Quantile* | *−b* | *ln(a)* | *a* |
| Orientale (660) | -0.95 ± 0.17 | 9.77 ± 1.21 | $1.8 \times 10^4$ |
| Copernicus (93) | -0.68 ± 0.06 | 4.82 ± 0.33 | $1.2 \times 10^2$ |
| Kepler (31) | -0.33 ± 0.10 | 1.62 ± 0.44 | 5.1 |
| Unnamed in SPA (3.0) | -0.39 ± 0.13 | -1.22 ± 0.39 | $2.9 \times 10^{-1}$ |
| Unnamed near Orientale (2.2) | -0.05 ± 0.03 | -2.59 ± 0.11 | $7.5 \times 10^{-2}$ |
| Unnamed in Procellarum (0.83) | 0.10 ± 0.06 | -3.61 ± 0.14 | $2.7 \times 10^{-2}$ |
| *99.9th Quantile* | | | |
| Copernicus (93) | -0.60 ± 0.08 | 4.75 ± 0.46 | $1.2 \times 10^2$ |

\*$d_{sec,max} = aR^{-b}$, where R is in km, units of a are $km^{b+1}$. Note: All parameters for the 99th quantile were used in Fig. 9. The estimates are reported as the parameter ± the standard error ($1\sigma$; see section 3).

**Table 3.** *Generalized Fits to all Six Secondary Crater Fields as shown by Black Lines in Fig. 8*

| Function[a] | Regression parameters[b] | | |
|---|---|---|---|
| *Secondary Craters: $d_{sec,max} = aR^{-b}$* | | | |
| $a = CD^m$ | $m = 1.96 \pm 0.15$ | $ln(C) = -3.82 \pm 0.56$ | $C = 0.022$ |
| $-b = n\ln(D) - E$ | $n = -0.14 \pm 0.03$ | | $E = 0.00 \pm 0.09$ |
| | | | |
| *Ejecta Fragments: $d_{frag,max} = \alpha v_{ej}^{-\beta}$* | | | |
| $\alpha = FD^p$ | $p = 3.91 \pm 0.39$ | $ln(F) = 3.2 \pm 1.40$ | $F = 24.8$ |
| $-\beta = r\ln(D) - G$ | $r = -0.40 \pm 0.05$ | | $G = 0.28 \pm 0.18$ |

[a]R (range) and D (primary crater diameter) are in km, and $v_{ej}$ (ejecta fragment velocity) in m s$^{-1}$. See additional details in Sections 3 and 4.

[b]Linear fits to the natural log of the data were run to determine the equations for a and α. C, m, E, and n are the regression parameters for each of the four fits to a, α, −b, and −β as functions of D (see Fig. 9 for curves plotted along with data points). The estimates are reported as the parameter ± the standard error ($1\sigma$; see section 3).

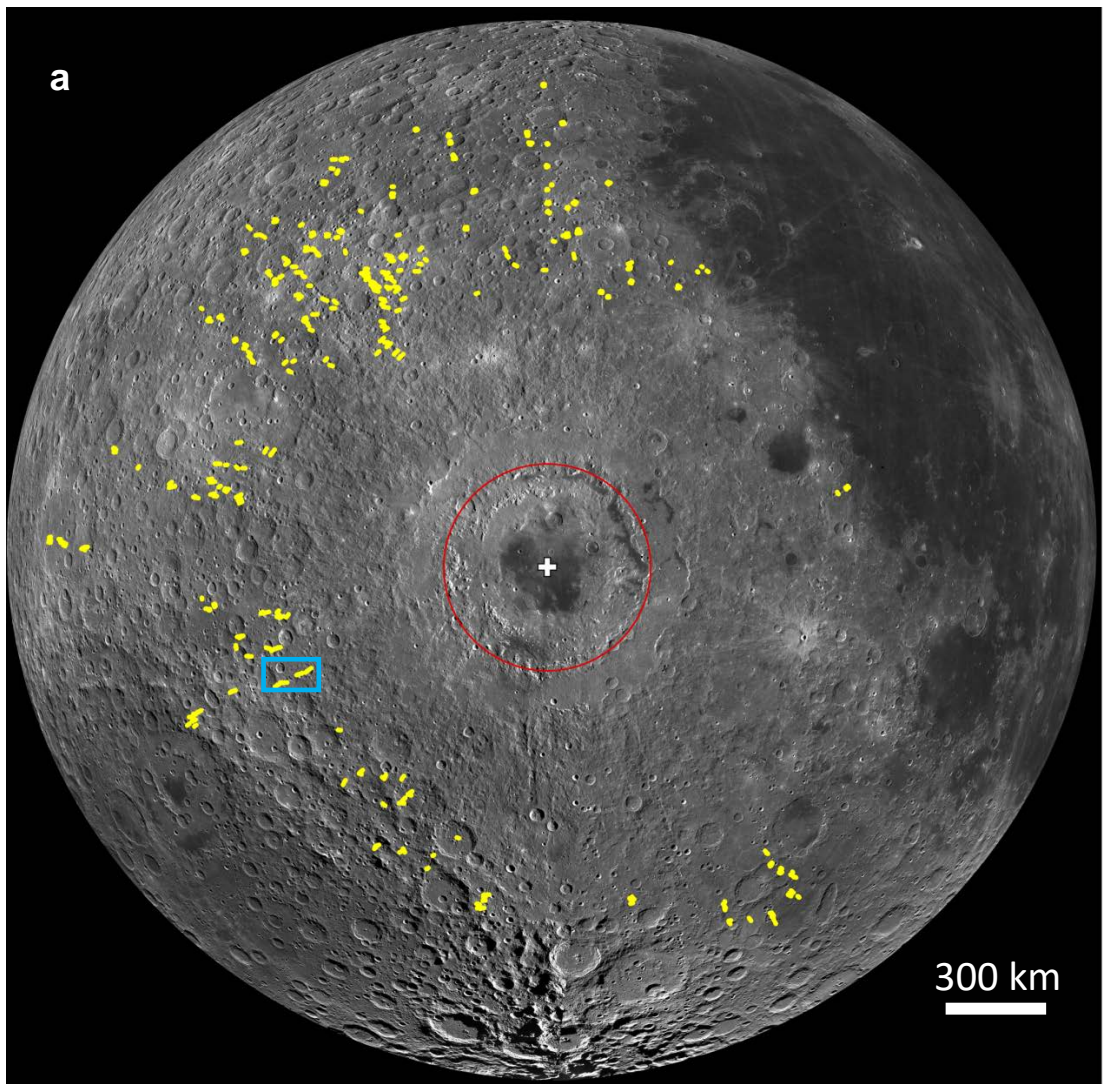

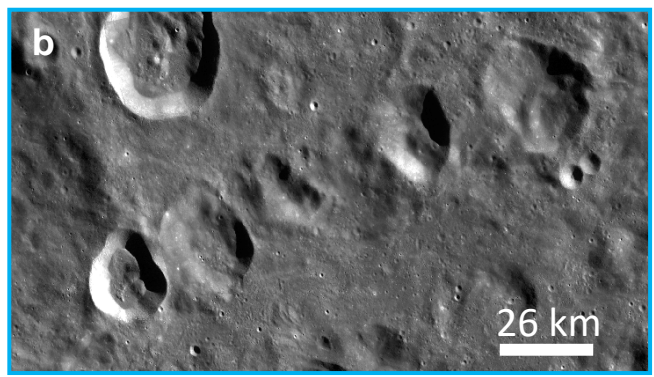

**Figure 3**. **Orientale basin. a,** 660 km in diameter (approximate rim location near the Outer Rook Ring shown with red outline); orthographic projection centered at 19.1°S, 266.0°E.  Mapping conducted on WAC 100 m px$^{-1}$ basemap.  **b,** Example of a large secondary crater chain.

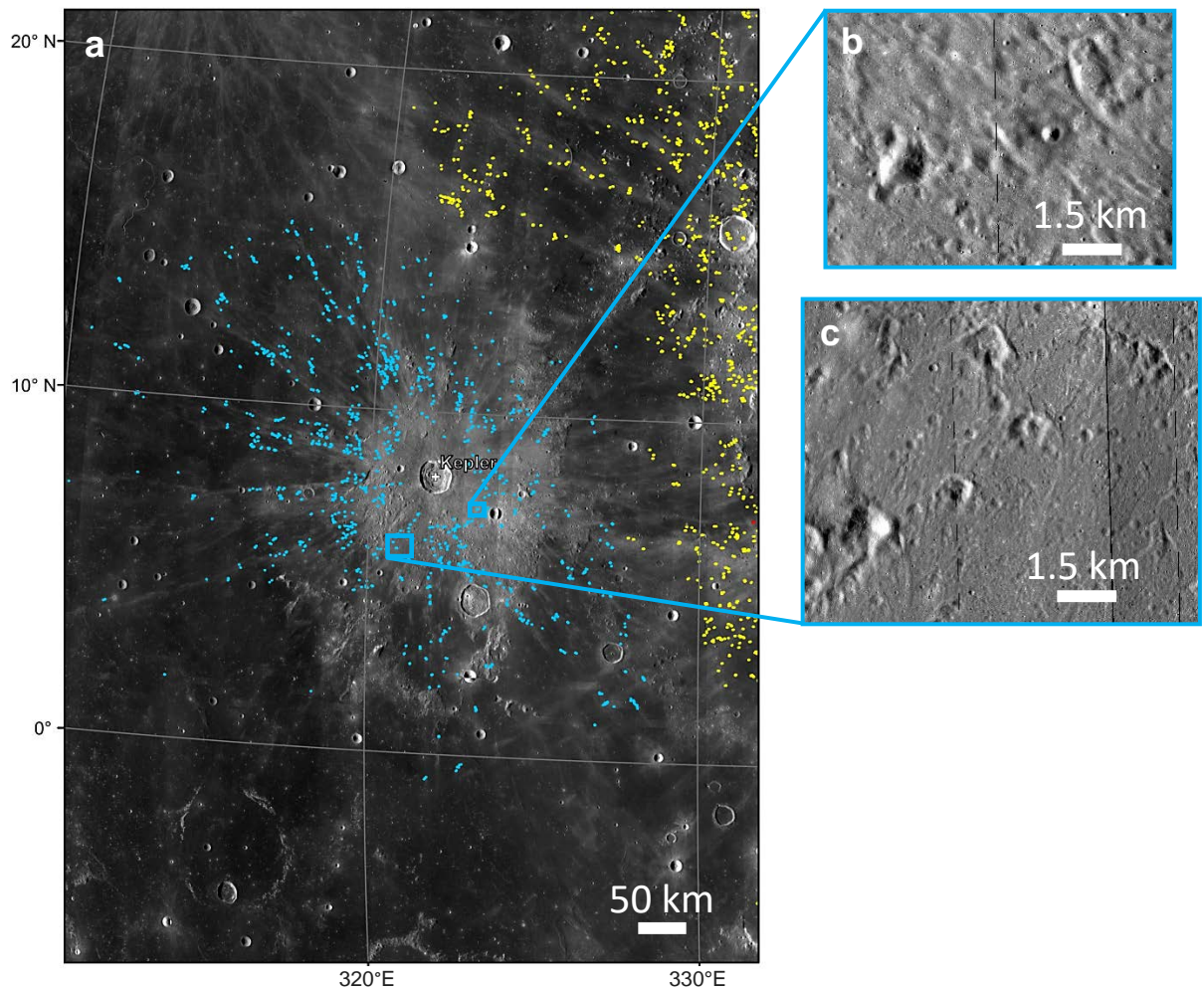

**Figure 4**. **Kepler crater. a,** 31 km in diameter; centered at 8.1°N, 322.0°E (orthographic projection is centered on Copernicus). The 100 m px⁻¹ WAC mosaic served as a basemap, but we also confirmed secondary crater morphologies in NAC images where available (using the LROC online QuickMap™ (ACT)). **b and c,** Examples of some of the larger secondaries.

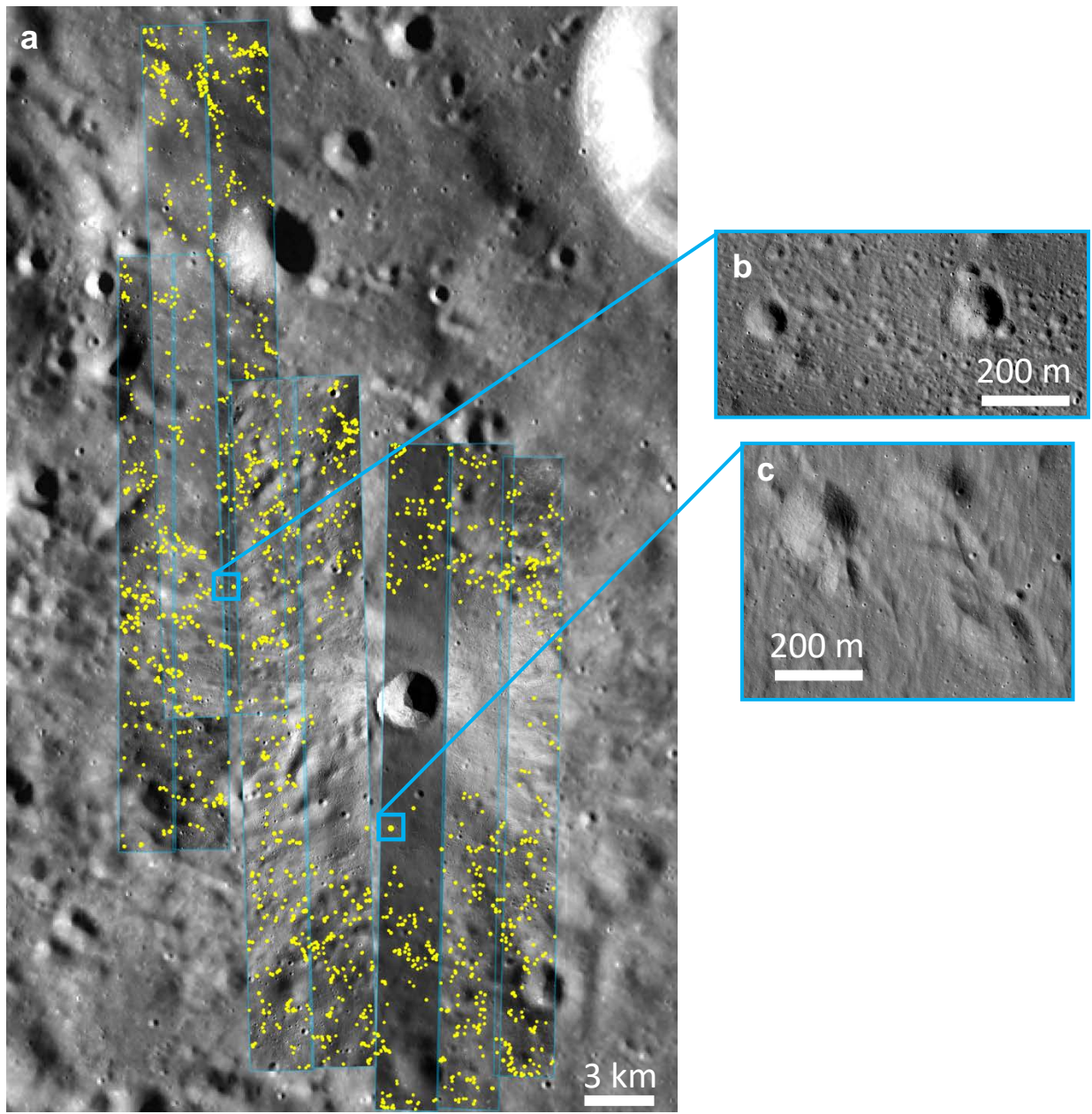

**Figure 5**. **Small primary in the South Pole Aitkin Basin. a,** 3.0 km in diameter; centered at 41.4°S, 188.2°E. Four NAC pairs plus one additional NAC were mapped (M161081870 L/R, M1107054125 L/R, M1097623854 L/R, M191723165 L/R, M112713912L). **b and c,** Examples of some of the larger secondaries.

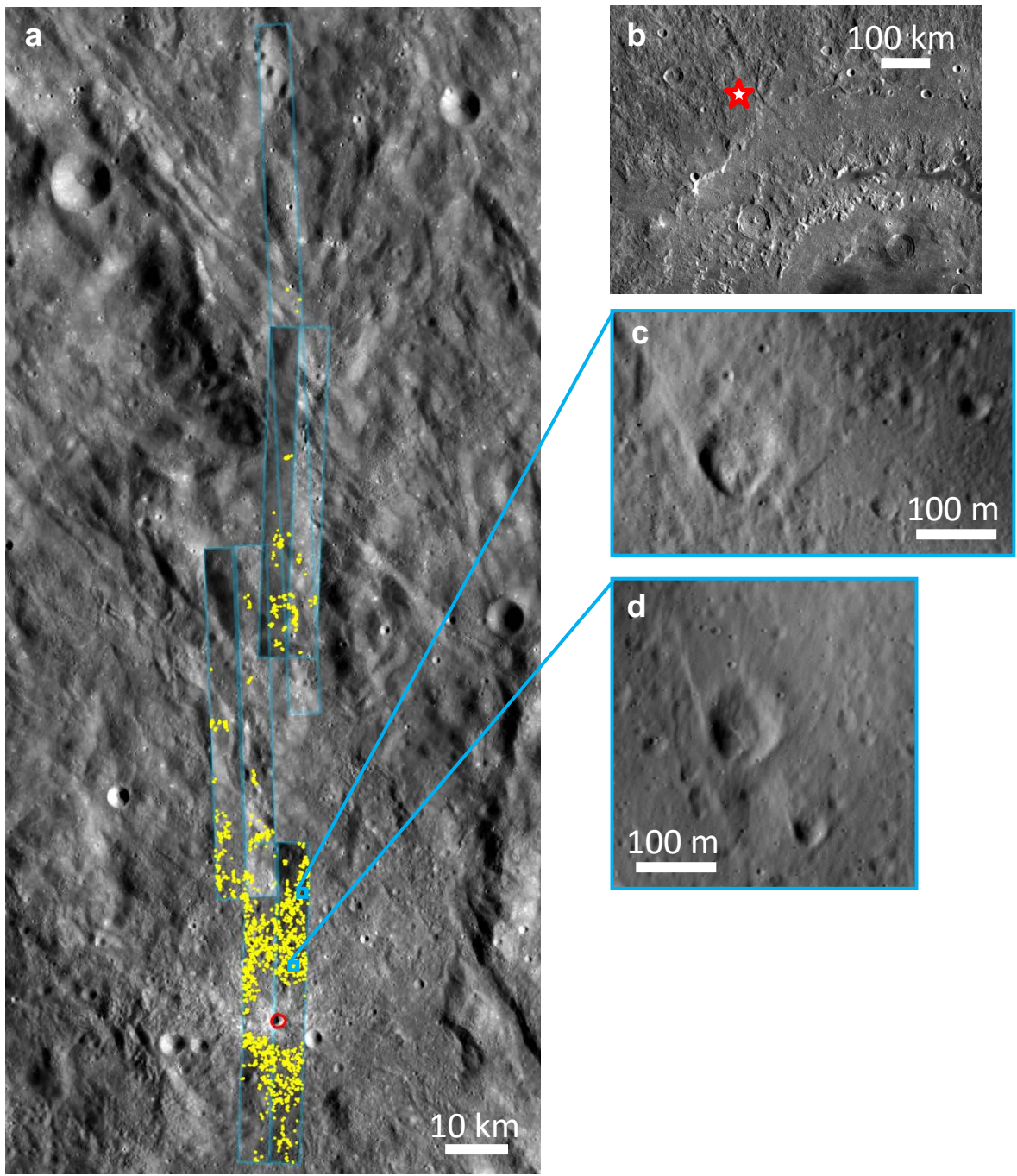

**Figure 6. Small primary in the proximal ejecta of Orientale. a,** 2.2 km in diameter; centered at 5.1°S, 255.6°E. Six NAC pairs plus one longer NAC (M1143115078 L/R, M1097180128 L/R, M1112494436 L/R, M1139592131R). **b,** Context image, star marks small primary location (WAC 100 m px⁻¹). **c and d,** Examples of some of the larger secondaries.

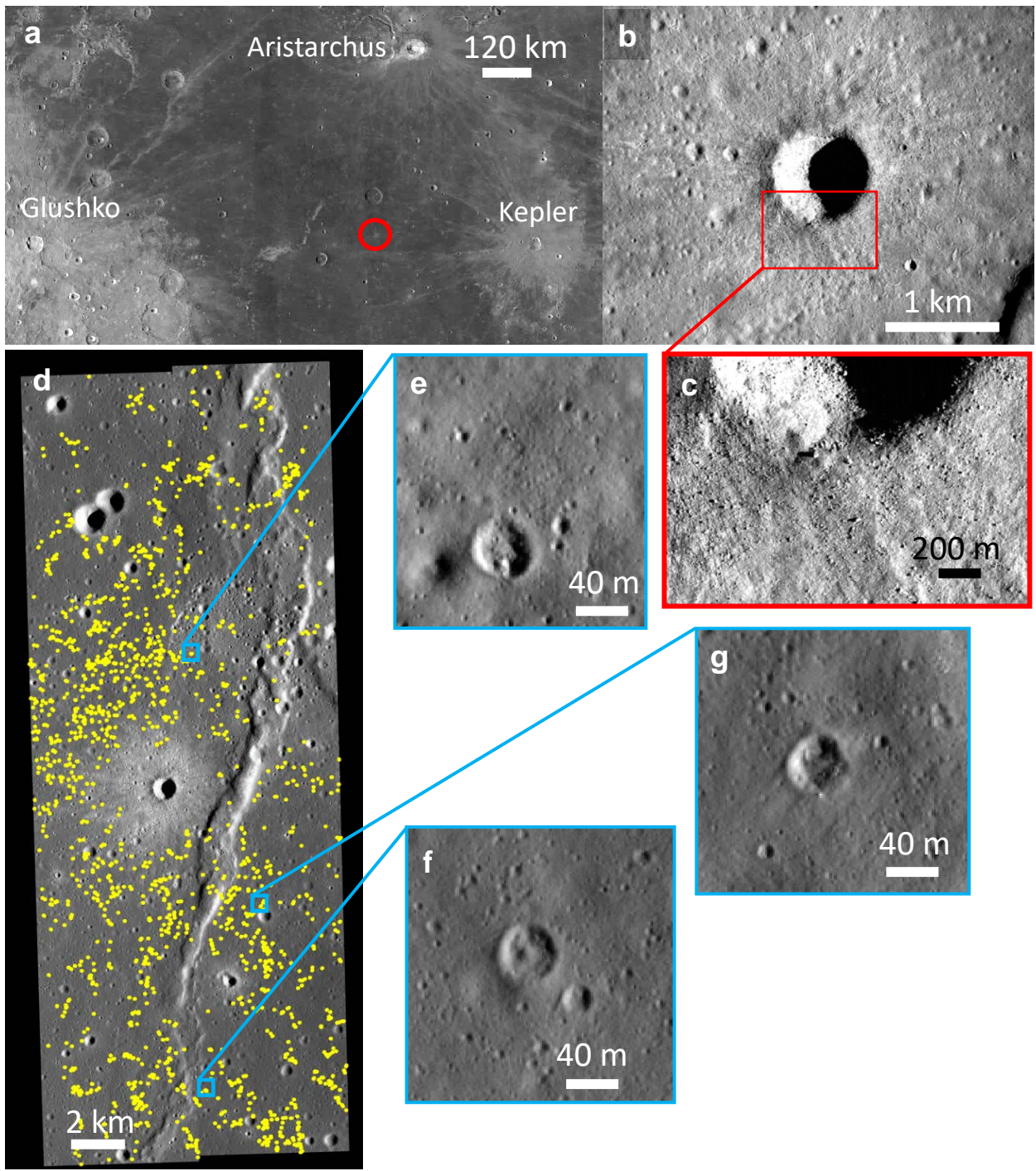

**Figure 7**. **Small primary in Oceanus Procellarum.** 0.825 km in diameter; centered at 8.7°N, 309.3°E. One NAC pair (M1108611064 L/R). **a,** Context images for this primary (WAC 100 m px$^{-1}$). **b and c,** Close up images of primary. **d,** Mapped secondaries. **e-g,** Examples of largest secondaries (40 m scale bar applies to all three).

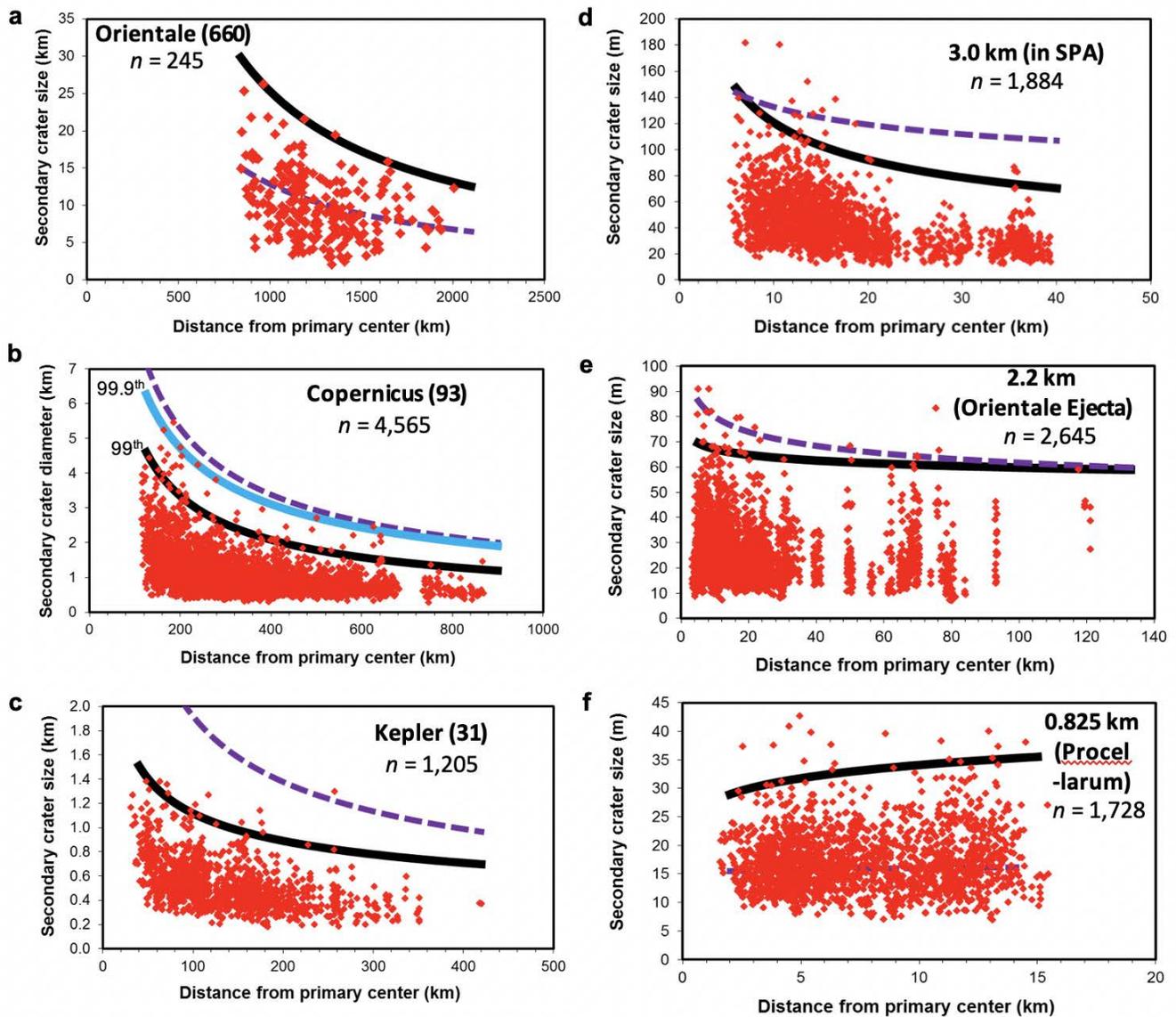

**Figure 8. All measured secondary sizes and ranges from the primary impact crater.** Black curve is the 99[th] quantile regression fit (Table 2). The 99.9[th] quantile fit is also shown in panel b with a blue line for reference. The purple, dashed line is the fit for the generalized equation given in Table 3. It can be seen that the generalized fit can be offset by as much as ~50% for some fields, thus we recommend using the individual fits provided in the supplement (Tables S2 and S3) for primary craters similar in diameter to those studied here. Gaps in the range coverage for a given secondary field (x-axis) are due to either the lack of distinct secondaries in this range, or the image coverage constraints for secondary fields mapped in LROC NACs.

## Secondary Crater Quantile Regression Parameters (MSRD)

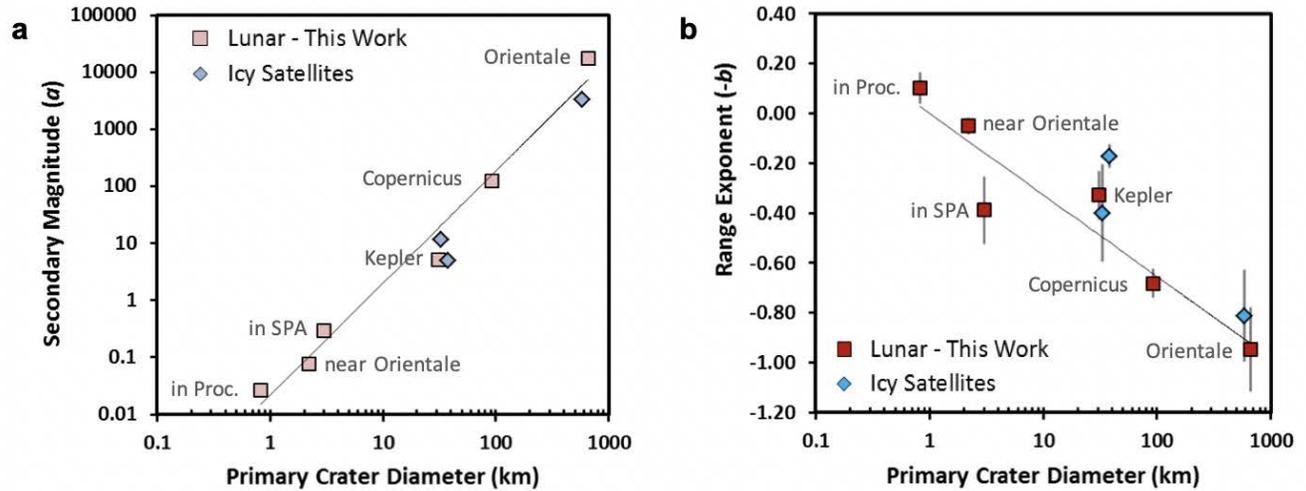

## Ejecta Fragment Quantile Regression Parameters (MSVD)

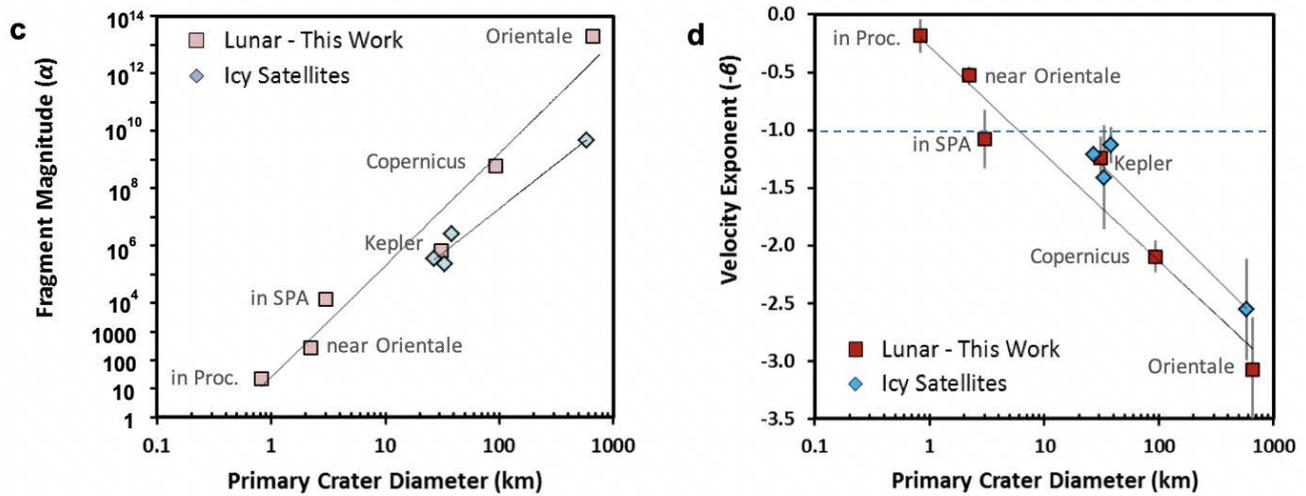

**Figure 9. Quantile regression parameters for all secondary crater fields.** Parameters shown are for the 99th quantile fit. Red squares indicate fit parameters for the six secondary fields in this work, blue triangles are for a similar study of three secondary fields on the icy satellites Europa and Ganymede (Singer et al., 2013). Equations for black fit lines are given in Table 3. **a,b** Parameters $a$ (units are km$^{b+1}$) and $-b$ for power law fit to upper envelope of each secondary crater size-range distribution (Table 2; see details in Section 3). **c,d** Parameters $\alpha$ (units are m$^{\beta+1}$ s$^{-\beta}$) and $-\beta$ for upper envelope of each estimated ejecta fragment size-velocity distribution (Table 4; see details in Section 4). The velocity exponent predicted by spallation theory (Melosh, 1984, 1989), $-\beta = -1$, is highlighted for reference in d.

The parameters in Table 2, or equations in Table 3 for interpolation between primary sizes, can be used to estimate the maximum secondary crater size at a given distance from a wide range of primary craters on the Moon. Natural variation and limited examples mean the Table 3 master equations under- or over-estimate the upper envelope for a specific crater, but are generally within an uncertainty of ± 50% (Fig. 8). Accordingly, we recommend using the individual quantile regressions (Table 2) for estimating MSRDs around primaries close in size to one of the six mapped. These results for secondary crater MSRDs can also be used in reverse, to constrain from which primary a secondary cluster or chain originated (Fig. 10). Using NAC images with similar lighting geometries, one can also confirm that secondaries close to the potential parent primary crater are similar in appearance/degradation state to more distant secondary clusters.

The largest secondaries are 4-6% of the primary size (Table 1, similar to previous results (Allen, 1979; Schultz and Singer, 1980; Shoemaker, 1965; Xiao et al., 2014)). The first distinctly identifiable secondaries begin at ~1.3 *D* from the rim (Fig. S1a) for the larger primaries (Orientale, Copernicus, and Kepler), with the largest secondaries occurring from ~1.3–2.5 *D*. For the smaller primaries (0.825–3 km in diameter), the first secondaries appear around 1.8 *D* and the largest occur over a larger range of scaled distances (*R/D*), as these smaller primary craters exhibit increasingly flat MSRDs (Fig. 8d-f and S1b).

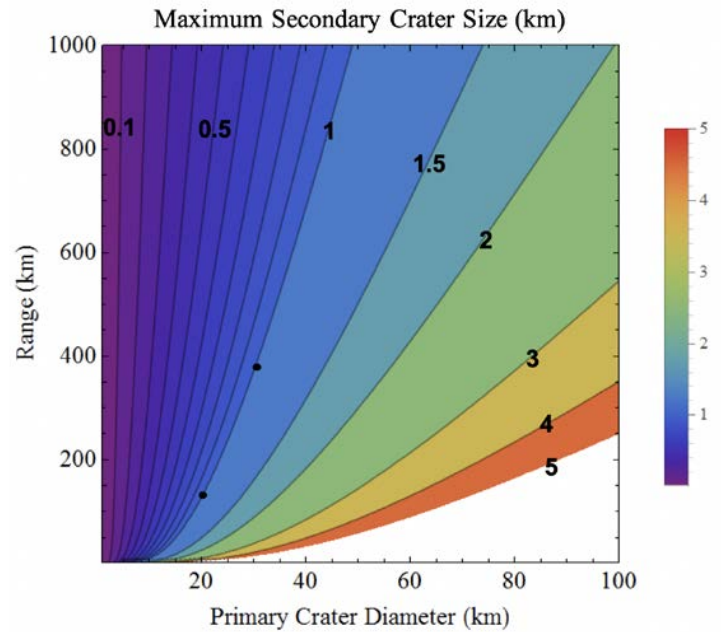

**Figure 10. Predicted maximum secondary crater sizes.** Secondary crater diameters as a function of both primary crater diameter (*D*) and distance from that primary (range or *R*) as estimated by the generalized fit equation in Table 3. Example points show how a 1 km secondary could be the result of a 20 km (or larger) primary crater 130 km away, or a 30 km primary 370 km away. Combined with observations of radial indicators, this analysis can narrow down the options for the parent crater of a given secondary or secondary chain/cluster. Natural variation in cratering (as displayed by the departure of the individually fitted quantile regression parameters from the general trend lines in Fig. 9 and Table 3) leads to suggested error bars of ± 50% for a number read off of this plot (also see Fig. 9).

## 4. Ejecta fragment maximum size-velocity distributions (MSVDs)

Studies with empirical, observation components such as this one complement and can be used to validate experiments and modeling, and can test theories of ejecta formation during impact crater excavation. Here we summarize the scaling used, and expand on the details for interested readers in the subsections below and in the supplement (and general forms are given in section S2). We estimate the velocity ($v_{frag}$) of the ejecta fragment that formed each secondary crater from the range equation for a ballistic trajectory on a sphere (Melosh, 1989). We estimate the size ($d_{frag}$) of the fragments from the Schmidt-Holsapple-Housen scaling relations (methods as in (Singer et al., 2013) and described in the supplement section S2). Scaling laws relate projectile size to the resulting transient crater diameter ($D_{tr}$; e.g., Holsapple, 1993)). The ejection and subsequent re-impact velocities and angles ($\theta$) of ejecta fragments are assumed to be the same on the Moon, as there should be minimal atmospheric effects (the fragments sizes considered here should not be greatly affected by any transient impact-generated plumes). With an estimate of the secondary crater volume (assuming a depth-to-diameter $H/D = 0.125$ and a paraboloid of revolution shape), secondary fragment velocity, and material parameters for both hard rock and regolith as endmembers, the following analytical expressions estimate fragment size in two regimes: for the gravity regime, on a hard rock (non-porous) target,

$$d_{frag} = 0.822 \, D_{sec}^{1.275} \left[ g / \left( v_{frag} \sin \theta \right)^2 \right]^{0.275}, \quad (1)$$

for the gravity regime, on a regolith (porous, sand-like) target,

$$d_{frag} = 0.756 \, D_{sec}^{1.205} \left[ g / \left( v_{frag} \sin \theta \right)^2 \right]^{0.205}, \quad (2)$$

for the strength regime, on a hard rock target,

$$d_{frag} = 6.790 \, D_{sec} / \left( v_{frag} \sin \theta \right)^{0.551}, \quad (3)$$

and for the strength regime, on a weaker regolith target

$$d_{frag} = 1.047 \, D_{sec} / \left( v_{frag} \sin \theta \right)^{0.410}, \quad (4)$$

where $g$ is surface gravity. The results for the derived ejecta fragment sizes for each of the six secondary crater fields are shown in Fig. 11.

In principle, these scaling laws apply to hypervelocity impacts with speeds of at least a few km s$^{-1}$, whereas ejecta fragments that form secondary craters are below the typical sound speeds of the surface materials in many cases. Regolith sound speeds may only be a few 100 m/s, however (Carrier et al., 1991). The secondaries measured here have derived velocities as low as 50 m s$^{-1}$ for the secondaries close to the smallest primary, and up to 1.4 km s$^{-1}$ for secondaries from Orientale. However, these scaling laws have been shown to be appropriate for some conditions outside of those of typical hypervelocity impacts even at low velocity (e.g., Gault and Wedekind, 1978; Hartmann, 1985; Holsapple and Schmidt, 1982; Ormö et al., 2015; Yamamoto et al., 2006). Application of scaling laws to secondary craters and the caveats involved are discussed in Appendix A of Singer et al. (2013). We also compare to boulder sizes in the proximal ejecta deposits and find a good agreement with the ejecta fragments sizes estimated for the secondaries near the primary (see section 4.4). The full solution spanning the gravity and strength regimes was solved numerically for scaling of all secondary craters, as some examples lie near the transition between regimes (see section S3 and Figs. S3-5 for the full function, parameters and plots with information on scaling of each secondary field). However, they could also be fairly well approximated by the equations above for the gravity regime (using equation 1 for secondaries around the three largest primaries, and equation 2 for secondaries around the three smaller primaries).

We discuss our choice of scaling assumptions for the secondary crater fields below.

### 4.1 Secondary crater depth-to-diameter ratio

Secondary craters are generally known to be shallower than primary craters. Topography and example profiles of Copernicus secondaries from a NAC Digital Elevation Model (DEM) are shown in Fig. 12. Primary craters on the Moon have canonical *H/D* values near 0.2 (e.g., Pike, 1974, 1977; Wood and Anderson, 1978), which is somewhat lower than some studies found for craters on the Earth (Grieve and Garvin, 1984). On the Moon few secondary craters have been directly measured, but one early study of secondary craters ranging from 200 m to 40 km in diameter found an average depth-to-diameter (*H/D*) of 0.11 for those with simple morphologies (Pike and Wilhelms, 1978). Measurements for 23 small craters (~250 to 950 m in diameter) identified as secondaries yielded *d/D* values of ~0.02–0.2 for degraded Copernicus secondaries (with most values clustered around 0.05), and ~0.07–0.14 for somewhat fresher Tycho secondaries (Basilevsky et al., 2018). Additional measurements are available for a few other worlds. Depth-to-diameter measurements of Europa's small craters (*D* < 3 km), thought to be predominantly secondary craters, range from ~0.05–0.22 with a peak or mode near 0.14 (Bierhaus and Schenk, 2010). A study of secondary craters from four primary craters on Mars found a similar range of values, from ~0.05–0.22 with a trend for higher velocity and larger secondary impacts to allow for deeper craters (although a wide range is still observed for these subsets) (Watters et al., 2017). For a subset of secondary craters with $D > 60$ m and a subset for $D > 100$ m, the median $H/D$ was found to be 0.112 and 0.127 respectively, and the mean was found to be 0.116 and 0.130, respectively (Watters et al., 2017).

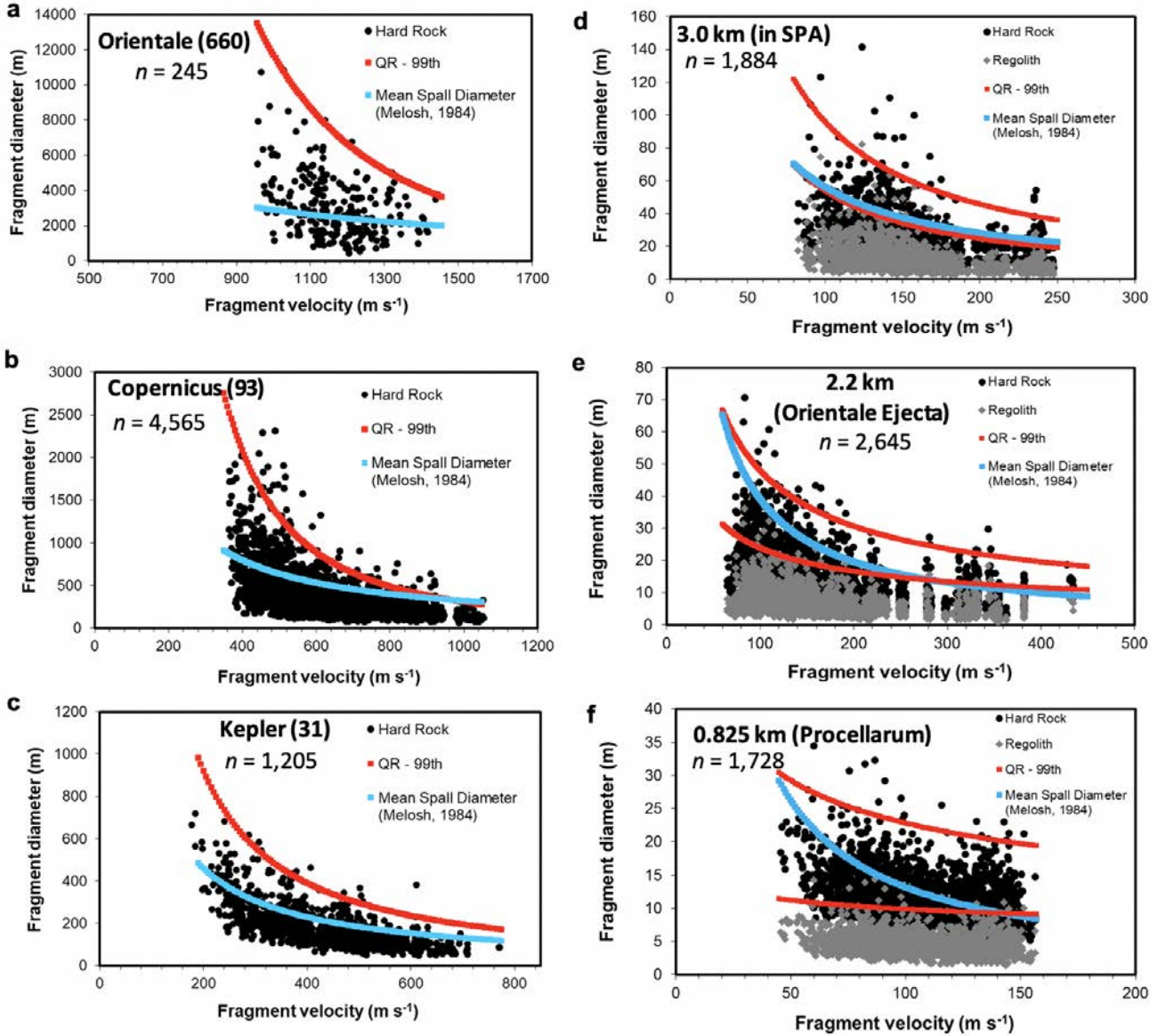

**Figure 11. All estimated fragment sizes and velocities.** The upper envelope red curve represents quantile regression fits (99[th] quantile, Table 4) for both hard rock (panels a-f) and regolith (panels d-f) scaling parameters. For reference, the blue curves give estimated mean spall diameters (see details in (Melosh, 1984)) calculated for a 15 km s[-1] primary impact speed (other parameters given in Tables 1 and S3). See discussion in Section 4 for scaling from secondary craters to ejecta fragment sizes. Both hard rock and regolith scaling is shown for the three smaller primaries as endmember materials for reference, though we argue that regolith scaling is the more appropriate in these cases (see text).

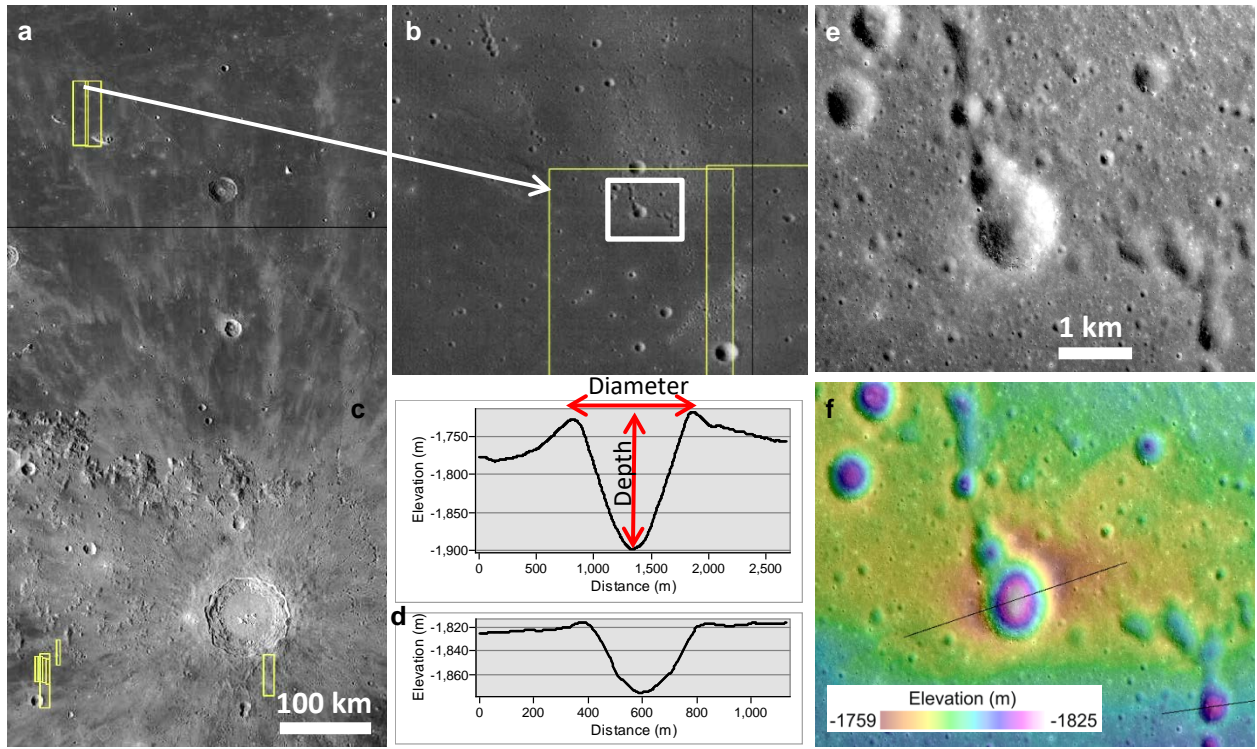

**Figure 12. Example secondary topographic profiles. a,** Context image showing location of available NAC DEMs near Copernicus (WAC 100 m px⁻¹ global mosaic as viewed from LROC QuickMap on 5-7-2014). **b,** Zoom showing location of secondary craters sampled here. **c and d,** Topographic profiles of two secondary craters (profile locations shown in **f**). The larger secondary crater (*D* = 1.09 km) has a rim-to-rim/rim-to-floor *H/D* of 0.16, and for the smaller (*D* = 465 m) *H/D* = 0.12. **e,** NAC image of Copernicus secondaries (M183697099, 1.5 m px⁻¹). **f,** NAC DEM of secondaries and location of example profiles (created from M183697099 L/R and M183711393 L/R, 5 m px⁻¹). These secondary depths cannot be measured in the global WAC DEM; NAC DEMs are necessary.

There have been some studies of the topography of the general population of small craters on the Moon. Many were aimed at studying primary craters, but could also include non-obvious secondary craters. The range of $H/D$ of small lunar craters ($D = 20$–260 m) measured in LROC NAC DEMs, including both primary and potential secondary craters (though obvious secondaries were excluded in most cases) with a range of degradation states, was found to be ~0.05-to-0.22 with an average of 0.11 (Basilevsky et al., 2014; Daubar et al., 2014; Mahanti et al., 2014; Stopar et al., 2017). Similar results were found for two NAC DEMs. A detailed study specific to the Apollo 16 and 17 landing sites found a wide range of $H/D$ for small craters (~30–250 m in diameter) for different degradation states and on different geologic units (Mahanti et al., 2018). Overall, 99% of the small craters measured have $H/D$ values between 0.04 and 0.17. The freshest morphologic class of craters with sharper rims and rays have $H/D$ values between 0.13 and 0.17 with a mean of 0.15 (Mahanti et al., 2018). A study of ~850 fresh small raters on both mare and highlands surfaces find $H/D$ values from 0.05–0.25 with a mean and median of 0.13 for craters 20 m to ~200 m in diameter (Sun et al., 2018). It is unknown what fraction of the crater population in these studies are secondaries, and also degradation plays a role, but these studies all found similar $H/D$ values to the range of depths found for the large population of secondary craters on Europa, and most find a mean or median $H/D$ between 0.11 and 0.15.

The measurements reported in the literature cited above are for the final crater $H/D$ measured from the rim to the floor, and our scaling is based on the most physically meaningful measure, the transient crater volume. Some collapse is expected between the final and transient craters, but on these lower gravity bodies, with lower impact velocities, there may not be as large of a difference between the final and transient crater $H/D$s, although more work is needed in this area (Orientale secondaries may be an exception, given their greater similarity to primary craters). We use $H/D = 0.125$ in the scaling equations, which is consistent with the plethora of the measurements described above (which are rim-to-floor depths), and is closer to the apparent depth (i.e., those measured from the pre-impact surface rather than the rim), .and it is also the value used by *Vickery* (1986; Vickery, 1987) derived from geometric constraints, and thus aids in comparison with previous work (see supplement section S3).

There are uncertainties when using any general crater shape assumptions, as there are for any other parameter values used in the scaling equations. The value of 0.125 with three significant digits should not be interpreted as statistical precision or accuracy, but is used here as a practical reference point based on the empirical measurements, crater geometry, and past work on this topic. Because the final depths for secondary craters are typically shallower than those of primaries, a shallower transient or apparent crater depth as we use here is likely more appropriate for secondary craters.

Adopting a different $H/D$ assumption would adjust the estimated size of secondary-forming fragments. For example, assuming a deeper $H/D$ ($> 0.125$) would systematically increase the fragment sizes but would have no effect on the MSVD slope ($-\beta$). This effect can be seen below in section 4.6 where $H/D = 0.2$ was used for the same scaling analysis of primary craters.

Comparing equations (1) and (9) shows that the multiplicative factor in front increases from 0.882 to 1.004 for $H/D$ = 0.125 and $H/D$ = 0.2, respectively, but that all other aspects of the equations remain the same. For this example the fragment sizes would increase by ~14%. Equivalently adopting a range for $H/D$ of 0.125 ± 0.025 would imply a variation in secondary fragment size of ± 8–10%, depending on whether strength or gravity scaling applies. Perhaps more important is the possible dependence of secondary $H/D$ on ejecta fragment launch/impact velocity. Watters et al. (2017) find, in their study of martian secondaries, that $H/D$ may be significantly lower for $v$ < 1 km s⁻¹. This effect is not well quantified yet, but should be the subject of future work.

### 4.2 Ejecta fragment velocities

For a ballistic trajectory on an atmosphereless, non-rotating sphere, the range of a projectile (or ground distance between the points of launch and re-impact), is given by

$$Range = 2R_p \tan^{-1}\left( \frac{v_{ej}^2 \sin\theta \cos\theta}{R_p g - v_{ej}^2 \cos^2\theta} \right),$$
(5)

where $R_p$ is the radius of the planet or moon, $g$ is surface gravity, $v_{ej}$ is the ejection velocity (also assumed to be the fragment impact velocity here), and $\theta$ is the ejection/impact angle measured from the ground-plane (Melosh, 1989). This equation is utilized to calculate the velocities ($v_{ej} = v_{frag}$) necessary for scaling. We used a value of half the transient crater radius as an estimate of the fragment launch location (see discussion in supplemental section S2 and Singer et al. (2013)).

A scale-dependent lower limit presumably exists for the velocity that can form a crater on the lunar surface. Proximal secondary craters, in this case those directly outside of the ejecta blanket, around the small primary craters measured here provide information about low velocity ejecta fragments and crater formation in regolith. Ejecta fragments from the 0.8-km-diameter primary crater, with velocities as low as 50 m s⁻¹, formed secondary crater-like features (Fig. 7). This is not necessarily a hard limit, as secondary fields around even smaller primary craters appear to have depressions that look like secondary craters. At such low velocities one might expect some strength or other scaling control on the secondary crater morphology, as deeper regolith is highly compacted (Carrier et al., 1991; Colwell et al., 2007). Although we have not done a systematic study of morphology changes with decreasing size, the features around the smallest primary considered in this work are still generally crater-like (Fig. 7). Most are fairly circular in planform, though some are misshapen. There are no clear signs of the impacting fragments, but some lumps on the secondary crater floors could resemble morphologies seen in low-impact-velocity studies (Hartmann, 1985). An exhaustive study has not been conducted in the present work to locate the smallest primary crater on the Moon with identifiable secondary craters. LROC NAC resolution limits the study of secondary craters to those ~5-10 m in diameter and larger.

Upper limits on the velocity of intact fragments (as opposed to melted or vaporized material) ejected during an impact are not well constrained. Some fragments (which may be spall or fragments from the main excavation) clearly reach escape velocity or higher, and some of these end up as meteorites on Earth. This study did not conduct an extensive search for identifiable secondaries at the great distances (corresponding to high velocities close to the escape velocity). For measurements in this study, fragments from Copernicus have estimated velocities up to ~1 km s⁻¹, and this cutoff is due the difficulty of following secondary crater rays into the highlands terrain north of Mare Imbrium (the direction of the most easily identifiable bright ray). Tycho secondaries have been noted at distances corresponding to ejection velocities of 1.5 km

s$^{-1}$ (Lucchitta, 1977). Secondary craters from the martian primary crater Zunil are noted as "certain" at distances corresponding to velocities of 2 km s$^{-1}$ and "probable" secondary craters were made by fragments traveling at 3 km s$^{-1}$ (Preblich et al., 2007). This evidence suggests that the 1 km s$^{-1}$ velocity cutoff put forward as a possibility by Vickery (1986) was likely a product of the specific cases considered therein and the data available at the time.

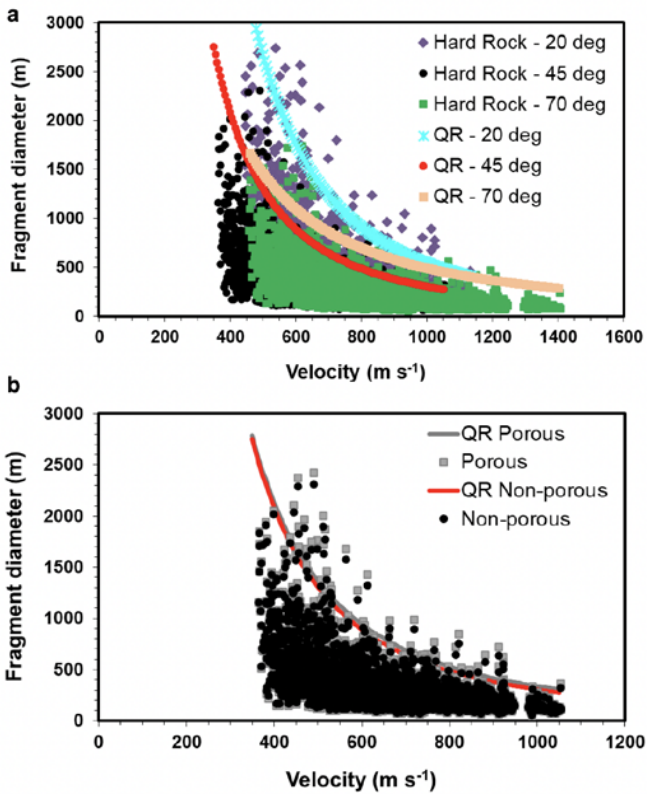

**Figure 13. Scaling parameter study.** Data for Copernicus showing **a,** the effect of changing the assumed ejection and impact angle of secondary fragments (the curves labelled QR give the quantile regression fits for each ejection angle case) and **b,** the effect of changing material parameter assumptions for gravity regime scaling. Calculated for plot b is in the gravity regime for two endmember material parameters: a porous material representing regolith (similar to dry sand but with a low strength), and non-porous hard rock. Porous

scaling yields fragments that are a few percent larger on average, although the effect is more pronounced for lower velocity fragments. We emphasize that these are endmember materials, to show a range of outcomes.

### 4.3 Ejection angle assumptions

The results shown here use $\theta = 45°$, because experiments show that most ejecta is launched at $\pm$ 15° from this angle (Cintala et al., 1999; Durda et al., 2012). Uniformly increasing the assumed angle to even 70° (very early ejecta generally leaves the crater at higher speeds and higher angles) only decreases estimated fragments sizes for Copernicus by ~25% (Fig. 13a). Some experiments indicate that ejection velocities are enhanced downrange of oblique impacts, and ejection angles may decrease (Anderson et al., 2003). Assuming an extremely low ejection angle of 20° increases the estimated fragment sizes by 35%. In both cases the overall velocity necessary to travel the same range increases, but the vertical component of the velocity is much smaller for the 20° ejection/impact angle, thus larger fragments are predicted by the crater scaling equations. For comparison (see Section 4.7 and Table 4), we have also calculated the quantile regression fits to the upper envelope of the MSVD for these more extreme ejection angle assumptions (Fig 13a) for the Copernicus secondary crater field. The MSVD power law values for the 20° case are: $-\beta = -2.29 \pm 0.18$, $\ln(\alpha) = 22.11 \pm 1.12$, and $\alpha = 4.01 \times 10^9$. The MSVD power law values for the 70° case are: $-\beta = -1.57 \pm 0.12$, $\ln(\alpha) = 17.04 \pm 0.72$, and $\alpha = 2.50 \times 10^7$.

### 4.4 Material parameter assumptions

Although the lunar surface is not a homogenous solid material, "hard rock" material-scaling parameters are likely more appropriate for the secondaries around the larger craters mapped here (see section S3 for full list of parameters). Even the smallest secondary craters mapped for Copernicus (~400 m) should penetrate any overlying regolith, which is estimated to be only a few meters thick on the nearside mare where the majority of Copernicus' secondary craters occur (Bart et al., 2011). For 3–10 meters of estimated regolith depth (Bart et al., 2011), and using an *H/D* of 0.125, secondary craters in the mare would need to be ~24–80 meters in diameter or smaller to be formed completely in regolith. The crust below this regolith would still be fractured, so the secondary fragment sizes calculated with hard rock material parameters should be considered a lower bound. For completeness we show the effect of using porous (which we sometimes refer to as regolith) vs. non-porous (hard rock) target properties in the gravity regime for Copernicus data in Fig. 13. We use these material parameters as approximate endmember cases, as the true material parameters likely lie somewhere in between. The porous/regolith material parameters are the same as that of sand in the gravity regime, but have a small yield strength and thus, as parametrized by the Schmidt-Holsapple-Housen scaling laws (described above), do change behavior for smaller impactors relative to that of dry sand, which has no cohesive strength (only friction). Some secondary craters may form from a clump of impacting fragments with a diameter equivalent to that predicted by scaling, which may help reconcile some of the large (>1 km) fragment sizes estimated (Fig. 11a,b) with the deeply fractured nature of the lunar crust (Wieczorek et al., 2013). Secondary craters up to ~150 m could form partially or wholly in regolith (i.e., 50% or more), thus regolith scaling is likely more appropriate for the secondaries around the three smaller primaries in Table 1.

Ejected boulders near the rim and in the ejecta blanket of a primary crater can provide an indirect comparison to ejecta fragments that create secondary craters. Many large boulders can be seen around the rim and on the ejecta blankets of the smallest three primaries. A rough measurement of a few of the largest boulders near the rim yields sizes of 20-30 m, 20-25 m, and 9-13 m for the 3.0-km-diameter, 2.2-km-diameter, and 0.8-km-diameter primaries, respectively. Blocks of this size are more consistent with the fragments sizes predicted from scaling with regolith material properties (as opposed to hard rock, see Fig. 11d,e,f). These block sizes scale with crater size, and so do not appear to be a preferred size inherited from pre-existing fractures/layering in the target (otherwise block sizes might be similar among the three primaries, at least on mare or impact melt units). We note that the basalt flow units in Oceanus Procellarum were estimated by deflections in crater size-frequency distributions there to be ~32–51 m thick (+10/-6 m; (Hiesinger et al., 2002)).

Moore (1971) analyzed boulders around large lunar craters, terrestrial explosion craters, and small experimental impacts and found a power-law relationship between the maximum block size and the primary crater size: $d_{block,max} \sim D^{2/3}$. Bart and Melosh (2010) found a similar relationship, $d_{block,max} = 0.40D^{0.65}$ (units are m), from boulder measurements around the rims of primary craters 0.23–41.2 km in diameter. Boulders and rock fragments measured by Cintala and McBride (1994) around the Surveyor sites were all less than 10 meters and data were not assigned to specific craters. Krishna and Kumar (2016) studied the boulder field around the bright-rayed, 3.8-km-diameter Censorinus crater and found large blocks near the rim were ~20–50 m in diameter, and maximum block sizes of ~10 m closer the edge of the ejecta blanket. *Watkins et al.* (2019) measured boulders around small primary craters near spacecraft landing sites ranging in diameter from 0.2 to 0.95 km. These craters (and boulder fields) had a range of degradation states but they report a linear relationship of $d_{block,max} = 0.022D$. Our measured block sizes are in accordance with these previous works. A discussion of general scaling motivated by these ejecta block studies can be found in Appendix C of Singer et al. Singer et al. (2013).

### 4.5 Scaling regime – strength vs gravity

The scaling regime for a given impact is determined by comparison of the gravity-scaled size ($\pi_2 = ga_i/U^2$) of the ejecta fragments to the theoretical transition from the strength to gravity regime for a given impact speed and target material (e.g., Holsapple, 1993). The range of gravity scaled sizes ($\pi_2$) for all six secondary crater fields are plotted in Fig. 14 (this is expanded on in the supplement and Figs. S3-5), illustrating how these values compare to the scaling regimes for both hard rock and regolith parameters. For secondary craters around the smaller three primaries we consider scaling in both the strength and gravity regimes. Fragments impacting a porous material (like sand or regolith) in the gravity regime have been shown to excavate a smaller crater than fragments impacting non-porous hard rock, likely because energy is taken up in collapsing the porosity. This is an effect seen in cratering experiments, leading to the empirically derived scaling parameters (e.g., Holsapple, 1993); http://keith.aa.washington.edu/craterdata/scaling/theory.pdf). Changing the material parameters within the strength regime has a different effect, where the same impactor striking a porous, lower-strength target may excavate a larger crater than into a stronger, less-porous material (a possible example is seen on the single-crater scale in van der Bogert et al. (2010)). This means a smaller fragment may form the same size crater in a porous target (see right-hand column of Fig. 11 and Fig. 13b). Of course, natural conditions are likely more complex than these idealized cases. Strength may be scale and/or strain-rate dependent. Even relatively competent surfaces on the Moon are likely fractured and porous to some degree. And the near-surface involved in lunar cratering can encompass a variety of mechanical units, from regolith, to lava flows, to fractured bedrock, and more.

**Table 4.** *Quantile regression fits to 99th quantile of ejecta fragment size-velocity distributions\*, for both hard rock (non-porous) and regolith (porous) targets.*

| Primary Crater (diameter in km) | Quantile Regression Parameters | | |
|---|---|---|---|
| | $-\beta$ | $ln(\alpha)$ | $\alpha$ |
| *Hard Rock (Non-Porous) Target* | | | |
| Orientale (660)[†] | -3.07 ± 0.46 | 30.61 ± 3.24 | $2.0 \times 10^{13}$ |
| Copernicus (93)[†] | -2.09 ± 0.14 | 20.19 ± 0.91 | $5.9 \times 10^{8}$ |
| Kepler (31)[†] | -1.24 ± 0.19 | 13.39 ± 1.10 | $6.5 \times 10^{5}$ |
| Unnamed in SPA (3.0) | -1.14 ± 0.28 | 9.22 ± 1.42 | $1.0 \times 10^{4}$ |
| Unnamed near Orientale (2.2) | -0.65 ± 0.07 | 6.86 ± 0.34 | $9.5 \times 10^{2}$ |
| Unnamed in Procellarum (0.83) | -0.36 ± 0.12 | 4.78 ± 0.59 | $1.2 \times 10^{2}$ |
| *Regolith (Porous) Target* | | | |
| Orientale (660) | -2.82 ± 0.45 | 28.94 ± 3.15 | $3.7 \times 10^{12}$ |
| Copernicus (93) | -2.01 ± 0.14 | 19.69 ± 0.92 | $3.6 \times 10^{8}$ |
| Kepler (31) | -1.17 ± 0.20 | 12.92 ± 1.20 | $4.1 \times 10^{5}$ |
| Unnamed in SPA (3.0)[†] | -1.07 ± 0.25 | 9.51 ± 1.28 | $1.4 \times 10^{4}$ |
| Unnamed near Orientale (2.2)[†] | -0.53 ± 0.08 | 5.60 ± 0.42 | $2.7 \times 10^{2}$ |
| Unnamed in Procellarum (0.83)[†] | -0.18 ± 0.15 | 3.13 ± 0.69 | $2.3 \times 10^{1}$ |

\*$d_{frag,max} = \alpha v_{ej}^{-\beta}$, where $\alpha$ and $v_{ej}$ are in $m^{\beta+1} s^{-\beta}$ and $m\ s^{-1}$. [†]Indicates preferred material parameter and values displayed in Fig. 9. The estimates are reported as the parameter ± the standard error ($1\sigma$; see section 3).

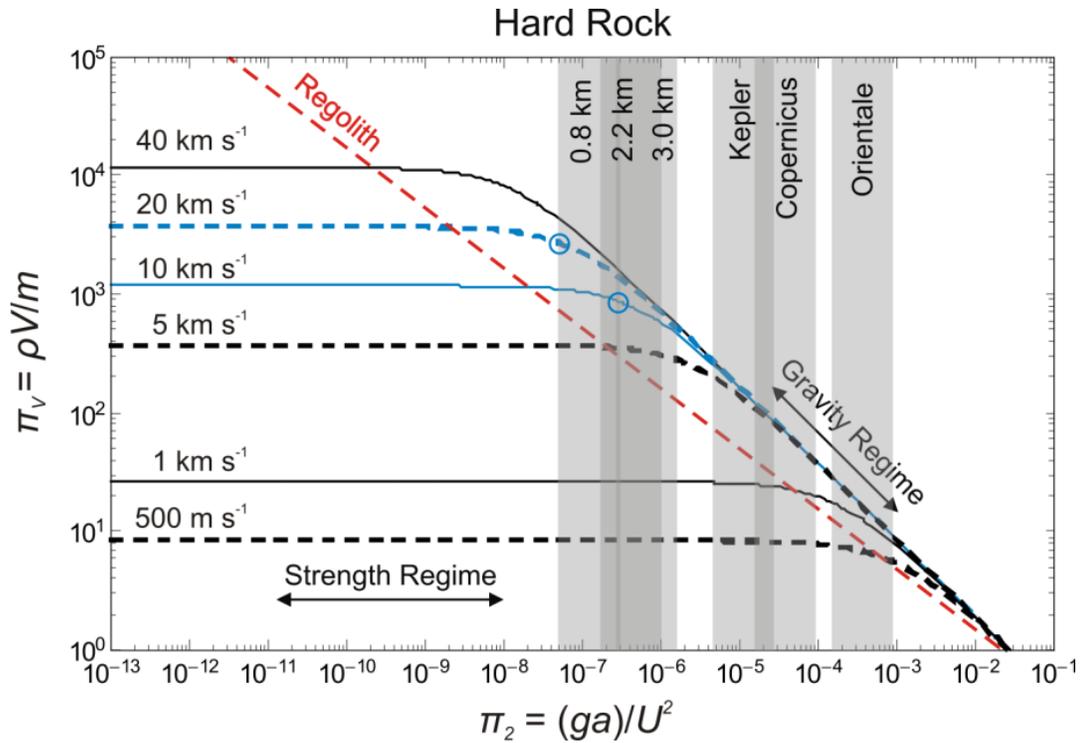

**Figure 14. Scaling for hard rock– comparison to primary crater gravity-scaled size.** Range of $\pi_2$ values shaded for each primary crater is for a primary impact speed of 20 km s$^{-1}$ (left bound) to 10 km s$^{-1}$ (right bound) – representing a typical range as estimated by modeling of asteroid impacts on the Moon in (Chyba, 1991), although the velocities could be higher. The very smallest primary ($D = 0.8$ km) is in a transitional zone between the strength and gravity regimes for this endmember material representing solid rock (illustrated with small blue circles). The two other small primaries ($D = 2.2$ and $3.0$ km) are near the transition, but in the gravity regime, and the three larger primaries are well into the gravity regime. All six primaries easily fall into the gravity regime for crater scaling in regolith (see supplement and Figs. S3-5 for similar details on the secondary crater scaling regimes).

*4.6 Calculations relating to the primary crater*

The transient diameter of the primary crater is used to determine the range or $R$ for equation (5). Transient cavity diameters for the larger, complex craters (Orientale, Copernicus, and Kepler) were estimated using the following equation, applicable to lunar craters:

$$D_{final} = k D_{tr}^{1.13}$$

$$k = 1.17\, D_c^{-0.13}$$

$$(6\text{-}8)$$

$$D_{tr} = 1.15 D_{final}^{0.885} \text{ (for the Moon)}$$

where $D_{final}$ is the final, measured crater rim-to-rim diameter, $D_c$ is the simple-to-complex transition diameter of 11 km for the Moon (McKinnon and Schenk, 1985; McKinnon et al., 2003), and all $D$ are in km. Simple power-law forms of this type have received some support from numerical models (Johnson et al., 2016). For the smaller, simple craters, the apparent diameter (diameter measured with respect to the level of the surrounding ground plane) was used as an approximation of the transient diameter. The apparent diameter was found by (Pike, 1977) to be on average ~0.83$D_{final}$ for simple craters; this result was used here.

Primary impactor size is used in the calculation of theoretical spall plate size (as shown in Fig. 2b and Fig. 11). Primary crater scaling also follows the Schmidt-Holsapple-Housen scaling laws, but for an assumed crater depth-to-diameter ratio (*H/D*) of 0.2 (the scaling laws refer to the transient, rather than the final crater size). The final equation for gravity regime, hard rock scaling is:

$$2a_i = 1.004 \, D_{tr}^{1.275}[g/(v_i \sin \theta)^2]^{0.275}$$
(9)

where $a_i$ is the primary impactor radius, $v_i$ is the primary impactor velocity, $g$ is gravitational acceleration, and $\theta$ is the impact angle (measured from the ground-plane), taken as 45° in all scaling calculations performed here, unless otherwise indicated. The $\sin\theta$ factor assumes that only the vertical component of the velocity is relevant for crater scaling, although this assumption may not be precisely true (Elbeshausen et al., 2009). Figures S3-5 illustrates this transition for hard rock material parameters and the $\pi_2$ for the primary craters are also plotted, which all fall into, or close to, the gravity regime. For the three smaller primaries, which are near the strength-to-gravity transition, scaling was carried out with a general equation that interpolates between the strength and gravity regime and solved numerically. The estimated impactor sizes were ~10 m different for each primary when scaled in this manner, compared with that for scaling in the gravity regime alone (equation 9). The Hevelius Formation of Orientale could be considered a weaker target material than hard rock. If the 2.2 km primary is scaled assuming lunar regolith parameters instead, as an endmember weak material, the predicted impactor diameter is 160 m (versus 114 m for hard rock).

*4.7 Fitting of ejecta fragment size-velocity distributions*

Quantile regression was used to fit a power-law function ($d_{frag,max} = \alpha v_{ej}^{-\beta}$) to the upper envelope (99th quantile) of the secondary fragment MSVDs (Fig. 11; Table 4). A power-law functional form is motivated by fragmentation theories (Melosh, 1984, 1989). Figure 9c-d and Table 4 present all derived fragment magnitudes factors ($\alpha$) and velocity exponents (-$\beta$), and the generalized expression as a function of $D$ and $v_{ej}$ is given in Table 3. For estimating fragment sizes from any given impact on the Moon, similar to the previous section, we recommend using the individual fits in Table 4 for any primary close in size to one measured here.

# 5. Discussion

*5.1 Scale-dependent trend in size-velocity distributions*

The maximum size of an ejected fragment falls off with increasing velocity and the rate of decline of fragment sizes is characterized by the velocity exponent parameter ($\beta$), as fit by the quantile regression described in the preceding paragraph. For reference, the spallation theory of *Melosh* (1984, 1989) predicts $\beta = 1$, and limiting $\beta$ values were derived for lunar craters [1,4/3] via coupling parameter scaling in the gravity regime, appropriate as long as the tensile strength is not rate-dependent (see discussion in (Singer et al., 2013) Appendix C). The spall ejection speed is also assumed to have no explicit dependence on the local sound speed in the target.

The $\beta$ parameter is observed, however, to vary between 0.2 and 3 on both icy and rocky bodies (Fig. 9; see also (Hirase et al., 2004; Hirata and Nakamura, 2006; Vickery, 1986; Vickery, 1987)). This unexpectedly large range in $\beta$ could mean a more complicated scaling applies than derived in Singer et al. (2013). For example, the "targets" may not be the same at large and small scales (e.g., solid rock vs. regolith), or the tensile strength, which governs spallation, may be rate dependent at larger scales. Regardless, the shallow $\beta$s found for the smaller primaries imply a weaker dependence on tensile strength than is exhibited by the larger craters with steeper $\beta$s (Singer et al., 2013, equation A.20). For the smaller craters, we may be seeing the influence of a fundamental block size controlled by preexisting structures such as faulting or regolith thickness; however, the estimated ejecta block sizes still scale with primary size even for the smallest primaries (as discussed in section 4.4). Wiggins et al. (2019) explored an implementation of Grady-Kipp fragmentation in iSale numerical modeling of vertical impactors 0.01–10 km in diameter and with velocities of 10–20 km s$^{-1}$. The results found only a weak fragment size dependence on impactor size and velocity for fragments in the deeper crust (fractured but not ejected material), but did find a trend for larger spall fragments with increasing impactor size.

Ejecta scaling by *Housen and Holsapple* (2011) described the total mass ejected at a given velocity with non-dimensional combinations of parameters that are important in the ejection process. These ideas motivate normalizing fragment velocities by $(g\ R_{tr})^{1/2}$ for gravity dominated impacts. Figure S2 illustrates how normalization brings the scaled fragment MSVDs for the three large craters into reasonable agreement. The individual $\beta$s, however, are not affected by this normalization. The scaled fragment sizes for Kepler and Copernicus are somewhat offset from each other, especially near the primary, where Kepler's scaled fragment sizes are smaller, giving its MSVD a shallower slope than that of Copernicus. The normalized sizes ($d_{frg}/R_{tr}$) of Orientale fragments are similar to those of Copernicus, thus there is no clear trend for larger primaries to have larger normalized fragments landing near the primary (Fig. S2a).

The data collected here presents a new scale-dependent trend in fragment sizes not predicted by any standing theories of fragmentation during impact events. The mechanical reasons for this trend may be revealed through future high-resolution numerical simulations.

*5.2 Relation to auto- or self-secondary cratering*

A growing body of evidence suggests some small craters ($D <$ a few hundred meters) on the ejecta blankets and melt-ponds of primary craters could be the result of ejecta from the primary itself (e.g., Artemieva and Zanetti, 2016; Plescia et al., 2010; Williams et al., 2014; Zanetti et al., 2017), rather than being later primaries or secondaries from other impacts. Shoemaker et al. (1969) were the first (to our knowledge) to propose self-secondaries in their study of the craters on the Tycho ejecta blanket around the Surveyor VII landing site. These super-proximal secondary craters would need to form after most of the ejecta blanket was emplaced, and thus would be delayed in the time sequence. If these secondaries form directly from ejecta in a manner similar to more typical secondary craters, the ejecta must be launched at a relatively high angle in order to land in the ejecta blanket, and to leave some time for the ejecta blanket to form. Very early ejecta could be higher angle than the bulk of the excavation flow, but it remains to be determined if any processes can separate high velocity, very high angle ejecta from the more typical high velocity, typical-angle (closer to 45°) ejecta from the main excavation flow that creates secondary craters.

Here we use the size-velocity distribution derived in this paper to consider the above scenario for self-secondary formation (recognizing that this is only one possible scenario). Using the 40-km-diameter Aristarchus crater as an example, at the highest possible velocity, just below escape velocity, an angle of 89.75° is necessary to land 0.25 $D$ (10 km) from the rim. The maximum time-of-flight for this high-velocity, high-angle ejecta would be ~40 minutes. Potential self-secondary craters 10–200 m in diameter on Aristarchus' ejecta blanket (Zanetti et al., 2017) imply ejecta fragments ~1–25 m. These fragment sizes are consistent with the sizes estimated at escape velocity from the MSVD analysis in this paper (maximum diameters in the 10s of meters for this scale of primary impact). This estimate represents an upper limit ejection velocity (also an upper limit approximate re-impact velocity), as it is possible that lower velocity fragments could also create self-secondaries. Another consideration is that some or even much of the earliest, highest velocity ejecta can be melted or vaporized (e.g., Johnson and Melosh, 2014; Melosh, 1989); thus if more competent fragments are required to produce self-secondaries they may not come from the very earliest ejecta stages of large primary impacts.

*5.3. Implications for lunar geologic mapping and meteorites*

Fragmentation and ejection during a planetary cratering event affects material exchange among planets, drives ejecta transport of material across a planet's surface, and has implications for the origin of small craters. Lunar secondary craters are abundant and can be quite large even at substantial distances from the primary. Additionally, both primary and secondary craters are heterogeneously distributed around the Moon. Thus, the size at which secondaries may contribute significantly to the crater population is not consistent across the Moon. Small craters (less than ~2 km in diameter) on the lunar nearside mare regions are dominated by secondaries from several mid-to-large-sized impacts (30-100 km in diameter), such as the thousands of secondary craters from Kepler and Copernicus as mapped in this work. The small crater distribution on the farside is a different mix of small primaries and secondaries from a myriad of larger primaries that each contributed their own scale-dependent distribution of secondary craters to the landscape. If absolute age dates from Apollo samples are tied to crater counts on the nearside, even if obvious secondary craters are accounted for in these calibrations, the size-frequency distribution of secondary craters in other locations on the Moon differs. This means that, at least locally, secondaries may be important even above the canonical 1 km diameter "cross-over" size, below which secondaries likely dominate the cratering record of the Moon (see comprehensive discussion in (McEwen and Bierhaus, 2006)). A similar effect is seen for martian craters (Robbins and Hynek, 2014) above a cross-over diameter of ~5 km. The size and location of secondaries can be roughly estimated with the results presented here, and potentially factored into interpretations based on crater counting.

Our results also provide constraints on the potential parent crater for a given secondary crater. For example, the secondary crater clusters thought to be from Tycho on the top of the South Massif south of the Apollo 17 site (Lucchitta, 1977, her Figure 7) are about 200-300 m and smaller in size. The 99$^{th}$ quantile fit parameters for Copernicus (93 km in diameter), as a proxy for Tycho (86 km), predict maximum secondary crater diameters up to 700 m at this distance (2250 km). This prediction easily allows that the South Massif secondaries, and many others on the Taurus-Littrow Valley floor, can be from Tycho (although some of the larger potential secondaries on the valley floor, ~700-800 m, have a more degraded appearance and could originate from other Copernican-aged craters). Many small craters, $D < 250$ m, near the Luna 23 and 24 sites in Mare Crisium, are thought to be secondaries from Giordano Bruno ($D = 22$ km), 1500 km away (Basilevsky and Head, 2012). Using 99$^{th}$ quantile parameters for the somewhat larger Kepler crater predicts maximum secondary cratersizes of 500 m at this distance from Giordano Bruno, thus secondaries a few hundred meters in diameter at the Luna 23/24 sites are plausible.

Extrapolating the power-law fits for maximum fragment size (Table 4) to the lunar Hill sphere escape velocity (2.34 km s$^{-1}$) yields escaping fragment sizes of a few meters for the smaller primary craters ($D < 3$ km), a few 10s of meters for mid-sized primary craters ($D = 30$–100 km), and up to ~900 m for the Orientale basin. At these high velocities, however, and given the fractured nature of the lunar regolith, megaregolith, or basement the material would likely break up in flight (e.g., Melosh, 1984). Thus, the estimates listed above and in Table 1 should be seen as upper limits, and certainly extrapolating from lower to higher velocity ejecta carries uncertainties, especially for the smaller primary craters considered here.

These high-velocity fragments would go into orbit about the Moon, Earth, or Sun, and may eventually return to impact the Moon (sesquinary cratering), or impact the Earth or other terrestrial planets as meteoroids. Dynamical modeling predicts that most lunar ejecta traveling to Earth on direct trajectories would do so on timescales of hours to days (Kreslavsky and Asphaug, 2014). Additionally, for ejecta speeds of 2.4 or 3.2 km s$^{-1}$, 50% or30% of fragments on heliocentric trajectories would impact a terrestrial body within ~1 million years, (Gladman et al., 1995). A particular ejecta fragment's dynamical evolution would depend on its launch position (latitude and longitude), ejection angle, and ejection velocity. Lunar meteoroid transit times, from their ejection to arrival at Earth, can be estimated from cosmic ray exposure ages; most transit times are 0.01–1 Ma, with the very longest estimated at a few-to-10s of millions of years (Jull, 2006). The youngest large crater Giordano Bruno ($D$ = 22 km), estimated to be less than 10 million years old (Morota et al., 2009; Plescia et al., 2010), could have ejected fragments as large as a few-to-10s of meters, later forming some of the lunar meteorite collection (Fritz, 2012). The median dynamical lifetime of the general population of near-earth objects, mainly originating from collisions in the asteroid belt, is estimated at ~10 million years (Gladman et al., 2000). Given the similarity of these timescales, it is possible that a small number of ejecta fragments from a lunar cratering event like Giordano Bruno could remain in near-Earth orbits.

## 6. Conclusions

We examined secondary crater fields around 6 primary craters on the Moon ranging in size from 0.83-to-660 km in diameter. We used the secondary crater diameters and their distances (or range) from their primary crater to estimate the size and velocity of fragments ejected during the primary crater formation through crater-to-impactor scaling laws. Only secondary craters in radial chains, clusters, or rays and with 2 or more morphological indicators that they were secondary craters were retained for the subsequent analysis.

We characterized the upper envelope of both the secondary crater size-range relationship and the fragment size-velocity distribution using quantile regression. This method yields an estimate of the maximum secondary crater size as a function of range (noted as MSRD in the text) and the maximum fragment size as a function of velocity (MSVD). The secondary crater sizes generally decrease with distance from the primary crater, as expected. However, the slope of the upper envelope (for either the secondary craters or the fragments) changes with primary size, where larger primaries have a steeper fall-off in sizes with distance, and the smallest primary examined has almost constant maximum secondary crater and fragment sizes with distance. Because this result appears in the secondary crater data themselves, this is not an effect of scaling to fragment sizes. This same overall trend (i.e., the steepness for the MSVD exponent with crater size) is found for secondary crater fields on the icy surface of Ganymede and Europa. This result suggests a scale-dependence in the fragmentation processes during the primary crater formation that has not previously been taken into account.

From the results in this paper we also compute a formula for estimating the maximum secondary crater sizes for a given distance away from any size primary crater on the Moon. This can be used to estimate the maximum size at which secondary craters may be a large part of the crater population at many locations on the Moon. We also provide a formula for estimating the size of fragments at a given velocity as a function of primary crater size. From the latter we estimate the maximum size of fragments that could potentially be ejected at escape velocity from an impact on the Moon, which ranges from a few meters up to 850 m for the primary craters studied here.

## Acknowledgments

We thank the LROC Team for many helpful discussions and comments on the paper, especially Alfred McEwen, Prasun Mahanti, Lilian Ostrach, and Julie Stopar. KNS also thanks Jeffrey Plescia, Randy Korotev, Axel Wittmann, Riccardo DiCecio, and Michael Zanetti for helpful discussions. We thank the JGR Associate Editor Gareth Collins, Natalia Artemieva, and Wesley Watters for very helpful comments and suggestions that improved this manuscript. The authors gratefully acknowledge funding for this work from NASA's LRO Project through subcontract NNG07EK00C from the University of Arizona (LROC) to Washington University (BLJ) and also from the NASA Solar System Working program grant NNX16AP97G to KNS. LROC image data are available through the Planetary Data Systems Geosciences node (https://pds-geosciences.wustl.edu/missions/lro/), and also on the LROC website (https://www.lroc.asu.edu/archive). Data files with all secondary crater diameters and locations are available at https://doi.org/10.6084/m9.figshare.11299319..

## References

Allen, C. C. (1979). Large lunar secondary craters - size-range relationships. *Geophysical Research Letters*, *6*, 51-54.

Anderson, J. L. B., P. H. Schultz, and J. T. Heineck (2003). Asymmetry of ejecta flow during oblique impacts using three-dimensional particle image velocimetry. *Journal of Geophysical Research (Planets)*, *108*, 5094.

Artemieva, N. A., and M. Zanetti (2016). Modeling small impact craters on ejecta blankets: Self-secondaries versus unrecognizable primaries. In *47th Lunar and Planetary Science Conference*, The Woodland, TX, Abstract #2143.

Arvidson, R., R. Drozd, E. Guinness, C. Hohenberg, C. Morgan, R. Morrison, and V. Oberbeck (1976). Cosmic ray exposure ages of Apollo 17 samples and the age of Tycho. In *Lunar and Planetary Science Conference Proceedings*, (Vol. 7, p. 2817-2832).

Bart, G. D., and H. J. Melosh (2010). Distributions of boulders ejected from lunar craters. *Icarus*, *209*, 337-357. https://doi.org/10.1016/j.icarus.2010.05.023

Bart, G. D., R. D. Nickerson, M. T. Lawder, and H. J. Melosh (2011). Global survey of lunar regolith depths from LROC images. *Icarus*, *215*, 485-490.

Basilevskii, A. T. (1976). On the evolution rate of small lunar craters. In *Lunar and Planetary Science Conference Proceedings*, (Vol. 7, p. 1005-1020).

Basilevsky, A. T., and J. W. Head (2012). Age of Giordano Bruno crater as deduced from the morphology of its secondaries at the Luna 24 landing site. *Planetary and Space Science*, *73*, 302-309.

Basilevsky, A. T., N. A. Kozlova, I. Y. Zavyalov, I. P. Karachevtseva, and M. A. Kreslavsky (2018). Morphometric studies of the Copernicus and Tycho secondary craters on the moon: Dependence of crater degradation rate on crater size. *Planetary and Space Science*, *162*, 31-40. https://doi.org/10.1016/j.pss.2017.06.001

Basilevsky, A. T., M. A. Kreslavsky, I. P. Karachevtseva, and E. N. Gusakova (2014). Morphometry of small impact craters in the Lunokhod-1 and Lunokhod-2 study areas. *Planetary and Space Science*, *92*, 77-87.

Bierhaus, E. B., C. R. Chapman, and W. J. Merline (2005). Secondary craters on Europa and implications for cratered surfaces. *Nature*, *437*, 1125-1127.

Bierhaus, E. B., A. S. McEwen, S. J. Robbins, K. N. Singer, L. Dones, M. R. Kirchoff, and J.-P. Williams (2018). Secondary craters and ejecta across the solar system: Populations and effects on impact-crater-based chronologies. *Meteoritics and Planetary Science*, *53*, 638.

Bierhaus, E. B., and P. M. Schenk (2010). Constraints on Europa's surface properties from primary and secondary crater morphology. *Journal of Geophysical Research*, *115*, 12004. https://doi.org/10.1029/2009JE003451

Carrier, W. D., G. R. Olhoeft, and W. Mendell (1991). Physical properties of the lunar surface, in *Lunar sourcebook*.

edited by G. Heiken et al., pp. 475-594, Cambridge University Press, Cambridge.

Chappelow, J. E. (2018). Detecting primary craters among clusters of secondaries. In *49th Lunar and Planetary Science Conference*, The Woodland, TX, Abstract

Chyba, C. F. (1991). Terrestrial mantle siderophiles and the lunar impact record. *Icarus*, *92*, 217-233.

Cintala, M. J., L. Berthoud, and F. Hörz (1999). Ejection-velocity distributions from impacts into coarse-grained sand. *Meteoritics and Planetary Science*, *34*, 605-623.

Cintala, M. J., and K. M. McBride (1994). Block distributions on the lunar surface: A comparison between measurements obtained from surface and orbital photography. In *Lunar and Planetary Science Conference*, (Vol. 25, p. 261-262).

Colwell, J. E., S. Batiste, M. Horányi, S. Robertson, and S. Sture (2007). Lunar surface: Dust dynamics and regolith mechanics. *Reviews of Geophysics*, *45*. https://doi.org/10.1029/2005RG000184

Daubar, I. J., C. Atwood-Stone, S. Byrne, A. S. McEwen, and P. S. Russell (2014). The morphology of small fresh craters on Mars and the Moon. *Journal of Geophysical Research (Planets)*, *119*, 2620-2639. https://doi.org/10.1002/2014JE004671

Denevi, B. W., M. S. Robinson, A. K. Boyd, H. Sato, B. W. Hapke, B. R. Hawke, and L. C. Cheek (2014). Crystalline, shocked, and melted materials in the lunar highlands. In *45th Lunar and Planetary Science Conference*, The Woodland, TX, Abstract #2000.

Durda, D. D., C. R. Chapman, W. J. Merline, and B. L. Enke (2012). Detecting crater ejecta-blanket boundaries and constraining source crater regions for boulder tracks and elongated secondary craters on Eros. *Meteoritics and Planetary Science*, *47*, 1087-1097. https://doi.org/10.1111/j.1945-5100.2012.01380.x

Elbeshausen, D., K. Wünnemann, and G. S. Collins (2009). Scaling of oblique impacts in frictional targets: Implications for crater size and formation mechanisms. *Icarus*, *204*, 716-731.

Fassett, C. I., and B. J. Thomson (2014). Crater degradation on the lunar maria: Topographic diffusion and the rate of erosion on the Moon. *Journal of Geophysical Research: Planets*, *119*(10), 2255-2271. https://doi.org/10.1002/2014je004698

Fielder, G. (1961). On the origin of lunar rays. *The Astrophysical Journal*, *134*, 425.

Fritz, J. (2012). Impact ejection of lunar meteorites and the age of Giordano Bruno. *Icarus*, *221*, 1183-1186.

Gault, D. E., and J. A. Wedekind (1978). Experimental impact "craters" formed in water: Gravity scaling realized. *EOS Trans. AGU*, *59*, 1121 (abstract).

Gladman, B., P. Michel, and C. Froeschlé (2000). The near-Earth object population. *Icarus*, *146*(1), 176-189. https://doi.org/10.1006/icar.2000.6391

Gladman, B. J., J. A. Burns, M. J. Duncan, and H. F. Levison (1995). The dynamical evolution of lunar impact ejecta. *Icarus*, *118*, 302-321.

Grady, D. E., and M. E. Kipp (1980). Continuum modelling of explosive fracture in oil shale. *International Journal of Rock Mechanics and Mining Sciences & Geomechanics Abstracts*, *17*(3), 147-157. https://doi.org/http://dx.doi.org/10.1016/0148-9062(80)91361-3

Grieve, R. A. F., and J. B. Garvin (1984). A geometric model for excavation and modification at terrestrial simple impact craters. *Journal of Geophysical Research*, *89*, 11561-11572.

Guest, J. E., and J. B. Murray (1971). A large scale surface pattern associated with the ejecta blanket and rays of Copernicus. *Moon*, *3*, 326-336.

Hartmann, W. K. (1985). Impact experiments 1. Ejecta velocity distributions and related results from regolith targets. *Icarus*, *63*, 69-98.

Hiesinger, H., J. W. Head, U. Wolf, R. Jaumann, and G. Neukum (2002). Lunar mare basalt flow units: Thicknesses determined from crater size-frequency distributions. *Geophysical Research Letters*, *29*, 1248. https://doi.org/10.1029/2002GL014847

Hirase, Y., A. M. Nakamura, and T. Michikami (2004). Ejecta size-velocity relation derived from the distribution of the secondary craters of kilometer-sized craters on Mars. *Planetary and Space Science*, *52*, 1103-1108.

Hirata, N., and A. M. Nakamura (2006). Secondary craters of Tycho: Size-frequency distributions and estimated fragment size-velocity relationships. *Journal of Geophysical Research*, *111*, E03005.

Holsapple, K. A. (1993). The scaling of impact processes in planetary sciences. *Annual Review of Earth and Planetary Sciences*, *21*, 333-373.

Holsapple, K. A., and K. R. Housen (2007). A crater and its ejecta: An interpretation of deep impact. *Icarus*, *187*, 345-

356.

Holsapple, K. A., and R. M. Schmidt (1982). On the scaling of crater dimensions 2. Impact processes. *Journal of Geophysical Research*, *87*, 1849-1870.

Honda, C., A. Shojyu, S. Suzuki, N. Hirata, T. Morota, H. Demura, M. Ohtake, J. Haruyama, and N. Asada (2012). Retention time of crater ray materials on the Moon. In *European Planetary Science Congress 2012*, (Vol. p. 806).

Housen, K. R., and K. A. Holsapple (2011). Ejecta from impact craters. *Icarus*, *211*, 856-875. https://doi.org/10.1016/j.icarus.2010.09.017

Johnson, B. C., G. S. Collins, D. A. Minton, T. J. Bowling, B. M. Simonson, and M. T. Zuber (2016). Spherule layers, crater scaling laws, and the population of ancient terrestrial impactors. *Icarus*, *271*, 350. https://doi.org/10.1016/j.icarus.2016.02.023

Johnson, B. C., and H. J. Melosh (2014). Formation of melt droplets, melt fragments, and accretionary impact lapilli during a hypervelocity impact. *Icarus*, *228*, 347-363. https://doi.org/10.1016/j.icarus.2010.09.017

Jull, A. J. T. (2006). Terrestrial ages of meteorites, in *Meteorites and the early Solar System II*. edited by D. S. Lauretta and H. Y. M. Jr., pp. 889-905, University of Arizona Press, Tucson.

Kenkmann, T., et al. (2018). Experimental impact cratering: A summary of the major results of the MEMIN research unit. *Meteoritics and Planetary Science*, *53*, 1543. https://doi.org/10.1111/maps.13048

Koenker, R. (2005). *Quantile regression*, Cambridge Univ. Press, New York.

Kreslavsky, M. A., and E. Asphaug (2014). Direct delivery of lunar impact ejecta to the Earth. In *45th Lunar and Planetary Science Conference*, The Woodland, TX, Abstract #2455.

Kreslavsky, M. A., J. W. Head, G. A. Neumann, M. A. Rosenburg, O. Aharonson, D. E. Smith, and M. T. Zuber (2013). Lunar topographic roughness maps from Lunar Orbiter Laser Altimeter (LOLA) data: Scale dependence and correlation with geologic features and units. *Icarus*, *226*, 52-66.

Krishna, N., and P. S. Kumar (2016). Impact spallation processes on the Moon: A case study from the size and shape analysis of ejecta boulders and secondary craters of Censorinus crater. *Icarus*, *264*, 274. https://doi.org/10.1016/j.icarus.2015.09.033

Kumar, S. P., K. J. Prasanna Lakshmi, N. Krishna, R. Menon, U. Sruthi, V. Keerthi, A. Senthil Kumar, D. Mysaiah, T. Seshunarayana, and M. K. Sen (2014). Impact fragmentation of Lonar Crater, India: Implications for impact cratering processes in basalt. *Journal of Geophysical Research: Planets*, *119*(9), 2029-2059. https://doi.org/10.1002/2013je004543

Lucchitta, B. K. (1977). Crater clusters and light mantle at the Apollo 17 site - a result of secondary impact from Tycho. *Icarus*, *30*, 80-96.

Mahanti, P., M. S. Robinson, and R. Stelling (2014). How deep and steep are small lunar craters? New insights from LROC NAC DEMs. In *45th Lunar and Planetary Science Conference*, The Woodland, TX, Abstract #1584.

Mahanti, P., M. S. Robinson, T. J. Thompson, and M. R. Henriksen (2018). Small lunar craters at the Apollo 16 and 17 landing sites - morphology and degradation. *Icarus*, *299*, 475-501. https://doi.org/10.1016/j.icarus.2017.08.018

Marcus, A. H. (1966). A stochastic model of the formation and survival of lunar craters v. Approximate diameter distribution of primary and secondary craters. *Icarus*, *5*, 590-605.

McEwen, A. S., and E. B. Bierhaus (2006). The importance of secondary cratering to age constraints on planetary surfaces. *Annual Review of Earth and Planetary Sciences*, *34*, 535-567.

McEwen, A. S., B. S. Preblich, E. P. Turtle, N. A. Artemieva, M. P. Golombek, M. Hurst, R. L. Kirk, D. M. Burr, and P. R. Christensen (2005). The rayed crater Zunil and interpretations of small impact craters on Mars. *Icarus*, *176*, 351-381.

McKinnon, W. B., and P. M. Schenk (1985). Ejecta blanket scaling on the Moon and - inferences for projectile populations. In *16th Lunar and Planetary Science Conference*, The Woodland, TX, Abstract #544-545.

McKinnon, W. B., P. M. Schenk, and J. M. Moore (2003). Goldilocks and the three complex crater scaling laws*Workshop on Impact Cratering*, Abstract #8047.

Melosh, H. J. (1984). Impact ejection, spallation, and the origin of meteorites. *Icarus*, *59*, 234-260.

Melosh, H. J. (1989). *Impact cratering: A geologic perspective*, Oxford Univ. Press, New York.

Moore, H. J. (1971). Large blocks around lunar craters, in *Analysis of Apollo 10 photography and visual observations*. edited, pp. 26-27, NASA Spec. Publ. SP-232.

Morota, T., et al. (2009). Formation age of the lunar crater Giordano Bruno. *Meteoritics and Planetary Science*, *44*, 1115-1120.

O'Keefe, J. D., and T. J. Ahrens (1985). Impact and explosion crater ejecta, fragment size, and velocity. *Icarus*, *62*, 328. https://doi.org/10.1016/0019-1035(85)90128-9

O'Keefe, J. D., and T. J. Ahrens (1987). The size distributions of fragments ejected at a given velocity from impact craters. *International Journal of Impact Engineering*, *5*, 493-499.

Ormö, J., I. Melero-Asensio, K. R. Housen, K. Wünnemann, D. Elbeshausen, and G. S. Collins (2015). Scaling and reproducibility of craters produced at the Experimental Projectile Impact Chamber (EPIC), Centro de Astrobiología, Spain. *Meteoritics & Planetary Science*, *50*(12), 2067-2086. https://doi.org/10.1111/maps.12560

Pike, R. J. (1974). Depth/diameter relations of fresh lunar craters: Revision from spacecraft data. *Geophysical Research Letters*, *1*, 291-294. https://doi.org/10.1029/GL001i007p00291

Pike, R. J. (1977). Apparent depth/apparent diameter relation for lunar craters. In *Lunar and Planetary Science Conference Proceedings*, (Vol. 8, p. 3427-3436).

Pike, R. J., and D. E. Wilhelms (1978). Secondary-impact craters on the Moon: Topographic form and geologic process. In *Lunar and Planetary Science Conference*, (Vol. 9, p. 907-909).

Plescia, J. B., M. S. Robinson, and D. A. Paige (2010). Giordano Bruno: The young and the restless. In *41th Lunar and Planetary Science Conference*, The Woodland, TX, Abstract #2038.

Preblich, B. S., A. S. McEwen, and D. M. Studer (2007). Mapping rays and secondary craters from the martian crater Zunil. *Journal of Geophysical Research (Planets)*, *112*, E05006.

Robbins, S. J., and B. M. Hynek (2014). The secondary crater population of Mars. *Earth and Planetary Science Letters*, *400*, 66-76. https://doi.org/10.1016/j.epsl.2014.05.005

Roberts, W. A. (1964). Secondary craters. *Icarus*, *3*, 348-364.

Robinson, M. S., et al. (2015). New crater on the Moon and a swarm of secondaries. *Icarus*, *252*, 229-235. https://doi.org/10.1016/j.icarus.2015.01.019

Schmedemann, N., T. Kneissl, A. Neesemann, G. Michael, R. J. Wagner, C. A. Raymond, and C. T. Russell (2014). The signature of secondary cratering on 4 Vesta and Tethys. In *45th Lunar and Planetary Science Conference*, The Woodland, TX, Abstract #1960.

Schmedemann, N., et al. (2017). How the distribution of impact ejecta may explain surface features on Ceres and saturnian satellites. *European Planetary Science Congress*, *11*.

Schultz, P. H., and D. E. Gault (1985). Clustered impacts - experiments and implications. *Journal of Geophysical Research*, *90*, 3701-3732.

Schultz, P. H., and J. Singer (1980). A comparison of secondary craters on the Moon, Mercury, and Mars. In *Lunar and Planetary Science Conference Proceedings*, (Vol. 11, p. 2243-2259).

Shoemaker, E. M. (1965). Preliminary analysis of the fine structure of the lunar surface in Mare Cognitum. In *The Nature of the Lunar Surface: Proceedings of the 1965 IAU-NASA symposium*, (Vol. p. 23-77).

Shoemaker, E. M., R. J. Hackman, and Z. K. Mikhailov (1962). Stratigraphic basis for a lunar time scale. In *The Moon*, (Vol. 14, p. 289-300).

Shuvalov, V. V., and N. A. Artemieva (2015). Craters made by severely fragmented asteroids. In *46th Lunar and Planetary Science Conference*, The Woodlands, TX, Abstract #1442.

Singer, K. N., W. B. McKinnon, and L. T. Nowicki (2013). Secondary craters from large impacts on Europa and Ganymede: Ejecta size–velocity distributions on icy worlds, and the scaling of ejected blocks. *Icarus*, *226*, 865-884.

Speyerer, E. J., R. Z. Povilaitis, M. S. Robinson, P. C. Thomas, and R. V. Wagner (2016). Quantifying crater production and regolith overturn on the Moon with temporal imaging. *Nature*, *538*, 215-218. https://doi.org/10.1038/nature19829

Speyerer, E. J., M. S. Robinson, B. W. Denevi, and L. S. Team (2011). Lunar reconnaissance orbiter camera global morphological map of the Moon. In *42th Lunar and Planetary Science Conference*, The Woodland, TX, Abstract #2387.

Stöffler, D., G. Ryder, B. A. Ivanov, N. A. Artemieva, M. J. Cintala, and R. A. F. Grieve (2006). Chapter 5. Cratering history and lunar chronology, in *New views of the Moon*. edited by B. L. Jolliff et al., pp. 519-596, Mineralogical Society of America, Chantilly, V. A.

Stopar, J. D., M. S. Robinson, O. S. Barnouin, A. S. McEwen, E. J. Speyerer, M. R. Henriksen, and S. S. Sutton (2017). Relative depths of simple craters and the nature of the lunar regolith. *Icarus*, *298*, 34-48. https://doi.org/10.1016/j.icarus.2017.05.022

Sun, S., Z. Yue, and K. Di (2018). Investigation of the depth and diameter relationship of subkilometer-diameter lunar craters. *Icarus*, *309*, 61-68. https://doi.org/10.1016/j.icarus.2018.02.031

van der Bogert, C. H., H. Hiesinger, A. S. McEwen, C. Dundas, V. Bray, M. S. Robinson, J. B. Plescia, D. Reiss, K. Klemm, and L. Team (2010). Discrepancies between crater size-frequency distributions on ejecta and impact melt pools at lunar craters: An effect of differing target properties? In *41th Lunar and Planetary Science Conference*, The Woodland, TX, Abstract #2165.

Vickery, A. M. (1986). Size-velocity distribution of large ejecta fragments. *Icarus*, *67*, 224-236.

Vickery, A. M. (1987). Variation in ejecta size with ejection velocity. *Geophysical Research Letters*, *14*, 726-729.

Walker, E. H. (1967). Statistics of impact crater accumulation on the lunar surface exposed to a distribution of impacting bodies. *Icarus*, *7*, 233-242.

Watkins, R. N., B. L. Jolliff, K. Mistick, C. Fogerty, S. J. Lawrence, K. N. Singer, and R. R. Ghent (2019). Boulder distributions around young, small lunar impact craters and implications for regolith production rates and landing site safety. *Journal of Geophysical Research*. https://doi.org/10.1029/2019JE005963

Watters, W. A., C. B. Hundal, A. Radford, G. S. Collins, and L. L. Tornabene (2017). Dependence of secondary crater characteristics on downrange distance: High-resolution morphometry and simulations. *Journal of Geophysical Research (Planets)*, *122*, 1773-1800. https://doi.org/10.1002/2017JE005295

Wells, K. S., D. B. Campbell, B. A. Campbell, and L. M. Carter (2010). Detection of small lunar secondary craters in circular polarization ratio radar images. *Journal of Geophysical Research (Planets)*, *115*, 6008.

Wieczorek, M. A., et al. (2013). The crust of the Moon as seen by GRAIL. *Science*, *339*(6120), 671-675. https://doi.org/10.1126/science.1231530

Wiggins, S. E., B. C. Johnson, T. J. Bowling, H. J. Melosh, and E. A. Silber (2019). Impact fragmentation and the development of the deep lunar megaregolith. *Journal of Geophysical Research (Planets)*, *124*, 941. https://doi.org/10.1029/2018JE005757

Wilhelms, D. E., V. R. Oberbeck, and H. R. Aggarwal (1978). Size-frequency distributions of primary and secondary lunar impact craters. In *9th Lunar and Planetary Science Conference Proceedings*, Abstract #3735-3762.

Wilhelms, D. E., J. F. with sections by McCauley, and N. J. Trask (1987). The geologic history of the Moon. *Report Rep. 1348*.

Williams, J.-P., D. A. Paige, J. B. Plescia, A. V. Pathare, and M. S. Robinson (2014). Crater size-frequency distributions on the ejecta of Giordano Bruno. In *45th Lunar and Planetary Science Conference*, The Woodlands, TX, Abstract #2882.

Wolfe, E. W. (1981). The geologic investigation of the Taurus-Littrow valley, Apollo 17 landing site. *Washington : U.S. Govt. Print. Off., 1981.*, *45*.

Wood, C. A., and L. Anderson (1978). New morphometric data for fresh lunar craters. In *Lunar and Planetary Science Conference Proceedings*, (Vol. 9, p. 3669-3689).

Xiao, Z., R. G. Strom, C. R. Chapman, J. W. Head, C. Klimczak, L. R. Ostrach, J. Helbert, and P. D'Incecco (2014). Comparisons of fresh complex impact craters on Mercury and the Moon: Implications for controlling factors in impact excavation processes. *Icarus*, *228*, 260-275.

Yamamoto, S., K. Wada, N. Okabe, and T. Matsui (2006). Transient crater growth in granular targets: An experimental study of low velocity impacts into glass sphere targets. *Icarus*, *183*, 215-224. https://doi.org/10.1016/j.icarus.2006.02.002

Zanetti, M., A. Stadermann, B. Jolliff, H. Hiesinger, C. H. van der Bogert, and J. Plescia (2017). Evidence for self-secondary cratering of Copernican-age continuous ejecta deposits on the Moon. *Icarus*, *298*, 64-77. https://doi.org/10.1016/j.icarus.2017.01.030

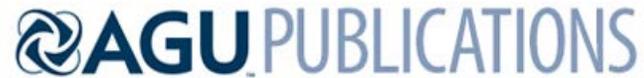

**Lunar Secondary Craters and Scaling to Ejected Blocks Reveals Scale-dependent Fragmentation Trend**

Kelsi N. Singer[1], Bradley L. Jolliff[2], and William B. McKinnon[2]

[1]Southwest Research Institute, 1050 Walnut St. Suite 300, Boulder, CO 80302,USA. [2]Department of Earth and Planetary Sciences and McDonnell Center for the Space Sciences, Washington University in St. Louis, Missouri 63130, USA.

## Contents of this file

Text S1 to S4
Figures S1 to S7
Table S1

## Additional Supporting Information (Files uploaded separately)

Caption for Table S2

## S1. Normalized data

Normalizing both the secondary diameters and their range values by the primary crater diameter ($D$) brings the secondary distributions of the larger three craters (Orientale, Copernicus, and Kepler) into fairly good agreement (Fig. S1a), illustrating the self-similarity of the cratering process (although the slopes for the upper envelopes of the individual distributions remain the same). Near the primary crater, Kepler's normalized secondary crater sizes are smaller than Copernicus', but larger than Orientale's ($d_{max,sec,Orientale}/D_{Orientale} < d_{max,sec,Kepler}/D_{Kepler} < d_{max,sec,Copernicus}/D_{Copernicus}$). It is possible the normalized fragments for Orientale are underestimated if the equivalent final rim-diameter ($D$) of Orientale is overestimated. The equivalent rim-diameter of Orientale is the least constrained, although the alignment of the left-most edge of the distribution in Fig. S1a indicates 660 km (Outer Rook) may be close to the correct size, given that the secondary field begins at the same scaled distance ($1.3D$) as for Copernicus and Kepler.

Alternatively, normalized Orientale secondary crater sizes could be misleading ($d_{max,sec,Orientale}/D_{Orientale}$) if the largest secondary craters around Orientale were not mapped for some reason. Although it formed close to 3.75 billion years ago (Stöffler et al., 2006), Orientale is the youngest major impact basin on the Moon, and many large secondary chains remain identifiable (Fig. 3). It is possible that some of the largest Orientale secondaries were assumed by us to be primary craters if they were not in a clear chain or cluster. It is more likely, however, that at least some of the largest secondaries from Orientale *are* in chains (as is true with Copernicus) and thus were mapped here. Mapping around more primaries may reveal if there is a trend in normalized fragments sizes with the size of the primary crater.

The 2.2-km crater has smaller normalized fragments ($d_{sec,max}/D$) impacting near the primary compared with the 3.0 km crater (Fig. S1c), similar to the relationship between Kepler (31 km) and Copernicus (93 km). The smallest crater (0.8 km in diameter), however, does not fit the same trend of decreasing normalized proximal secondary crater size (or ejecta fragment size, as shown in Fig. S2b) with decreasing primary size; its normalized fragment sizes are about the same as the 3.0 km primary. This 0.8 km primary is sufficiently small that fragmentation may be controlled more by the local regolith state (but see comment about block sizes in section 4.4). Further examples of secondary fields around small primaries should be sought to clarify any possible trends in normalized secondary crater or ejecta fragment sizes with primary diameter.

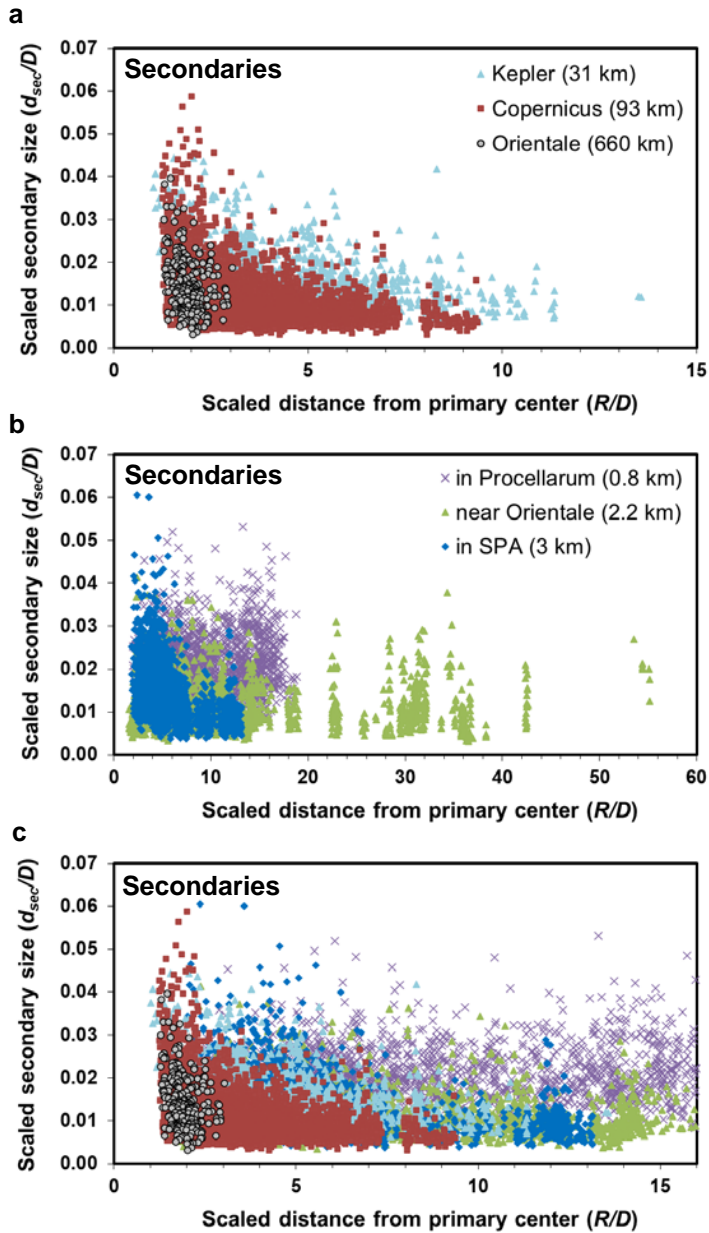

**Figure S1. Normalized secondary crater distributions. a,** Normalization by the final primary crater diameter (*D*; listed in parentheses for each primary) illustrates that distinct secondary craters first appear at similar scaled distances from the primary (*R/D*). Note that although the distributions align, the slopes for the upper envelopes on individual distributions are not affected by these normalizations. **b-c,** The smaller craters show increasingly flat distributions.

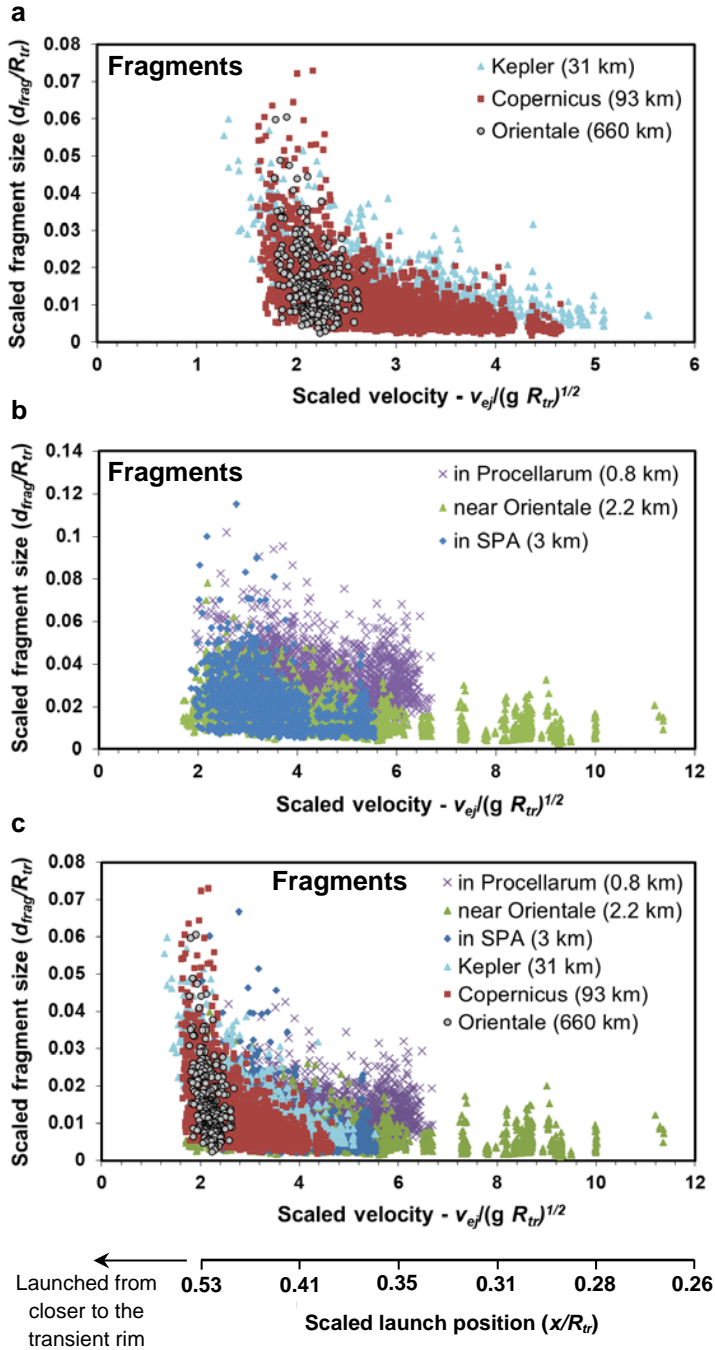

**Figure S2**. **Normalized ejecta fragment distributions.** **a,** The fragment sizes and velocities for the three largest primaries align when normalized by $R_{tr}$ and $(g \ R_{tr})^{1/2}$ (as motivated by (Housen and Holsapple, 2011), see text in this section), respectively. **b,** Normalized fragment sizes and velocities for the three smaller primaries. **c,** Comparison of estimated fragments for all six measured secondary crater fields. The scaled launch position are also shown in panel c for reference (see text in this section for details).

The scaled velocities ($v_{ej}/(g\,R_{tr})^{1/2}$) in this study have a somewhat wider range (~1.3–11; Fig. S2) than those of the secondaries into icy targets in Singer et al., 2013 (~1.6–4.8). The scaled velocities can be related to scaled launch positions ($x/R_{tr}$) where 0 is the center of the crater and 1 is the transient rim (e.g., Housen and Holsapple, 2011). The scaled launch position values relevant to this study are given in Fig. S2c for reference. As can be seen in Singer et al., (2013; their Fig. 13), which is modified from Housen and Holsapple (2011; their Fig. 14 and Eq. 14), the range of scaled velocities found in this study would suggest scaled launch positions of ~0.27 (for a scaled ejection velocity of 11) and ~0.62 (for a scaled ejection velocity of 1.3) for a moderate porosity material. The scaled launch position we used for our analysis is $0.5R_{tr}$ (or $0.25D_{tr}$) and equates to a scaled velocity of ~2.3. Similar to what is described in Singer et al., (2013), the effect of a different launch velocity on the distant secondaries in this study (those ejected from closer to the crater center with higher velocities) is minimal. The highest scaled launch velocities are for the 2.2-km-diameter primary (near Orientale) where secondaries were mapped at a large number of primary diameters away from the launch position, or about 55 km in range. Changing the 55 km range by $0.25R_{tr}$ (or $0.13D_{tr}$) has very little effect on our results. For Orientale, the scaled velocities are between ~1.7 and 2.7, yielding scaled launch positions between ~$0.56R_{tr}$ and ~$0.47R_{tr}$, respectively. These values are fairly close to our assumed value of ~$0.5R_{tr}$.

## S2. General Scaling Equation and Material Parameters

Equations 1-4 in the main text were derived from the general formulas for scaling (e.g., Holsapple, 1993; Housen and Holsapple, 2011 and references therein). Here we used the form that interpolates between the gravity and strength regimes as in *Holsapple* [1993; eqn 18]:

$$\pi_V = K_1\left\{\pi_2\left(\frac{\rho}{\delta}\right)^{(6v-2-\mu)/3\mu} + \left[K_2\pi_3\left(\frac{\rho}{\delta}\right)^{(6v-2)/3\mu}\right]^{(2+\mu)/2}\right\}^{-3\mu/(2+\mu)}$$

$$\pi_V = \frac{\rho V}{m}, \quad \pi_2 = \frac{ga}{U^2}, \quad \pi_3 = \frac{Y}{\rho U^2},$$

(S1)

where $K_1$ and $K_2$ are scaling coefficients for a given surface material, $\mu$ and $v$ are the scaling exponents, $\rho/\delta$ is the ratio of the target to impactor mass densities, $g$ is surface gravity, $V$ is the volume of the crater formed, $m$ is impactor mass, $a$ is impactor radius, $Y$ is a measure of the target strength, and $U$ is impact velocity (see **Table S1** below for values). The scaling parameters are empirically estimated for different materials from laboratory experiments, numerical computations, and comparison with explosion cratering results. The $\pi$-groups are non-dimensional: $\pi_V$ describes the overall cratering efficiency, and is dependent on the gravity-scaled size ($\pi_2$), and the strength measure ($\pi_3$). For the secondary craters considered here, we assume the density of the ejecta target fragments is similar to the density of the surface where it impacts, which reduces the equation to

$$\pi_V = K_1\left(\pi_2 + K_2\pi_3^2\right)^{\frac{-3\mu}{2+\mu}}.$$

(S2)

**Table S1.** *Scaling parameters*.

| | $K_1$ | $K_2$ | $\mu$ | $v$ | $Y$ (MPa) | $\rho$ (kg/m^3) |
|---|---|---|---|---|---|---|
| Water | 0.98 | 0 | 0.55 | 0.33 | 0 | 1,000 |
| Dry Sand | 0.132 | 0 | 0.41 | 0.33 | 0 | 1,700 |
| Wet Soil | 0.095 | 0.35 | 0.55 | 0.33 | 0.6 | 2,100 |
| Hard Rock | 0.095 | 0.257 | 0.55 | 0.33 | 10 | 3,200 |
| Lunar Regolith | 0.132 | 0.26 | 0.41 | 0.33 | 0.01 | 1,500 |

*These specific values are from a website maintained by Dr. Holsapple as downloaded in 2013 by the authors. The values are similar to those in *Holsapple* (1993), *Holsapple and Housen* (2007), and *Housen and Holsapple* (2011) . Note that wet soil and hard rock are the same for the gravity regime, and only differ in the strength regime. Additionally, dry sand and lunar regolith are also the same in the gravity regime, and the lunar regolith is modelled with a low strength whereas sand has zero strength.

In **Figures S3-5** we include a series of plots to illustrate the differences between scaling using different parameter sets representing different material properties. **Figure S3** compares the different parameter sets from Table S1. We chose to use the two endmembers of this group, which represent "regolith" or "hard rock" parameters, for the analysis in the main text to illustrate the range of possible outcomes. The exact parameters for the lunar crust are not rock known of course. We also discuss in the main text (and summarize in the caption of **Fig. S5**) which set of parameters are likely the most appropriate for each secondary field. **Figure S3** shows that, for the same velocity, an impact formed in regolith generally has a higher cratering efficiency in the strength regime (a smaller impactor is required to make the same size crater in regolith than in hard rock), but a lower cratering efficiency in the gravity regime, as compared to an impact into hard rock. **Fig. S4** illustrates the scaling for all of the secondary craters in the Copernicus secondary field would fall on a $\pi_2$ and $\pi_v$ plot for both hard rock and regolith scaling, and also the impact velocities for each secondary. **Figure S5** illustrates graphically where the $\pi_2$ and $\pi_v$ values for each secondary field would fall with respect to the strength and gravity regimes for both materials. For hard rock parameters, the secondaries for the larger three primaries fall primarily in the gravity or transition regimes. Note there is not a large difference in the estimated fragment diameter for secondaries falling into the transition regime vs purely the gravity regime. For the secondary fields around the smaller three primaries, scaling with hard rock parameters would put the secondary impacts into the strength regime, and would often suggest cratering efficiencies so low ($\pi_v$ below 1) that point-source-based scaling is no longer reliable. As described in the main text, the secondary crater morphologies we observe (depressions in the ground with no visible projectile) do not appear to be consistent with such very low cratering efficiencies. Nor are these morphologies consistent with the physical outcomes expected from very low velocity ($< 200$ m s$^{-1}$) impacts into hard igneous rock (cf. Kenkmann et al., 2018). For regolith material parameters, all secondary fields fall into the gravity regime.

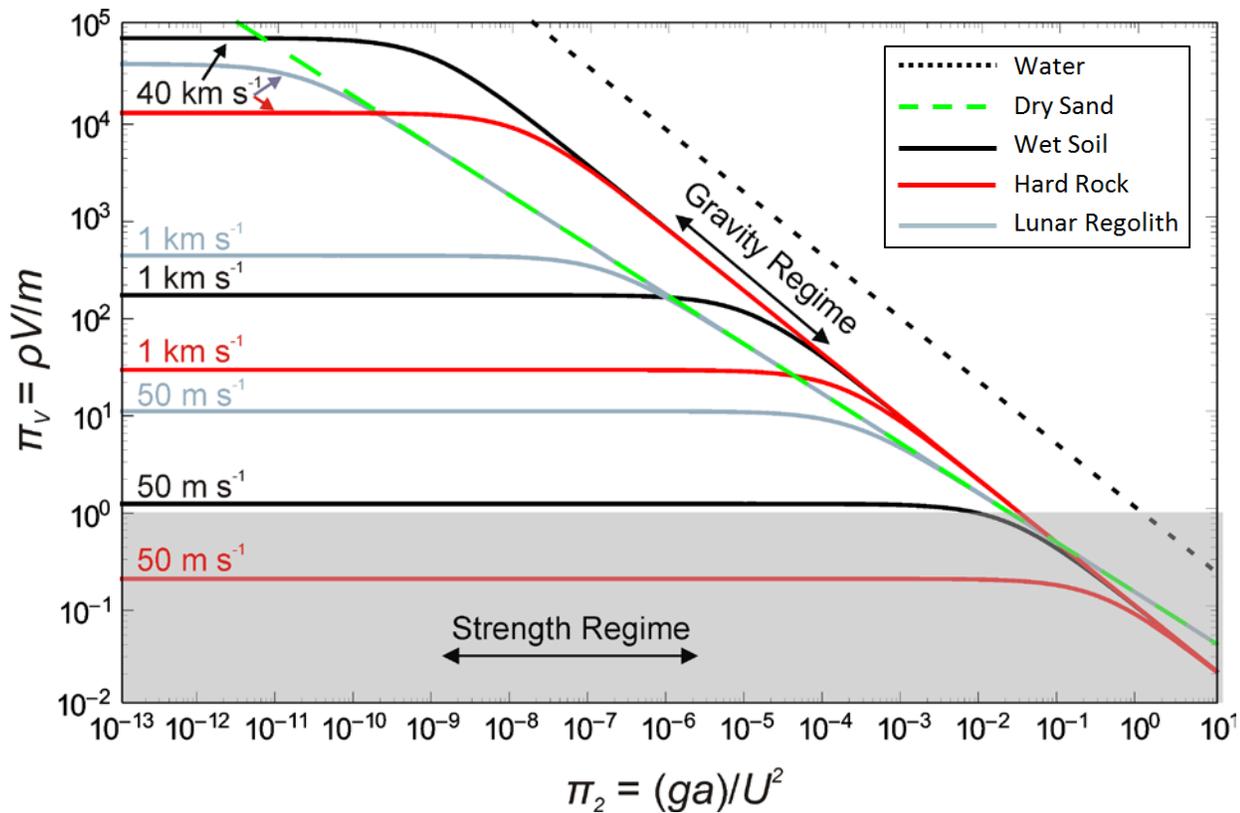

**Figure S3. Comparison of scaling with different material parameters.** The gravity scaled size ($\pi_2$) values and corresponding cratering efficiency ($\pi_v$) for different material parametrizations. See Table S1 for parameter list. For scaling in the gravity regime, the values would fall along the diagonal line labeled "Gravity Regime" at all velocities. For the strength regime, scaling would fall along the horizontal lines (for scale-independent strength), given for several velocities. For a given velocity the cratering efficiency (ratio of excavated mass to impactor mass) is modelled as constant for an increasing size of impactor, but in the gravity regime cratering efficiency decreases for increasing impactor sizes. The gray area in the lower portion of the Figure illustrates how low velocities lead to cratering efficiencies ($\pi_v$) less than ~one for hard-rock material parameters, which indicates that point-source scaling (the coupling parameter approach) is no longer valid.

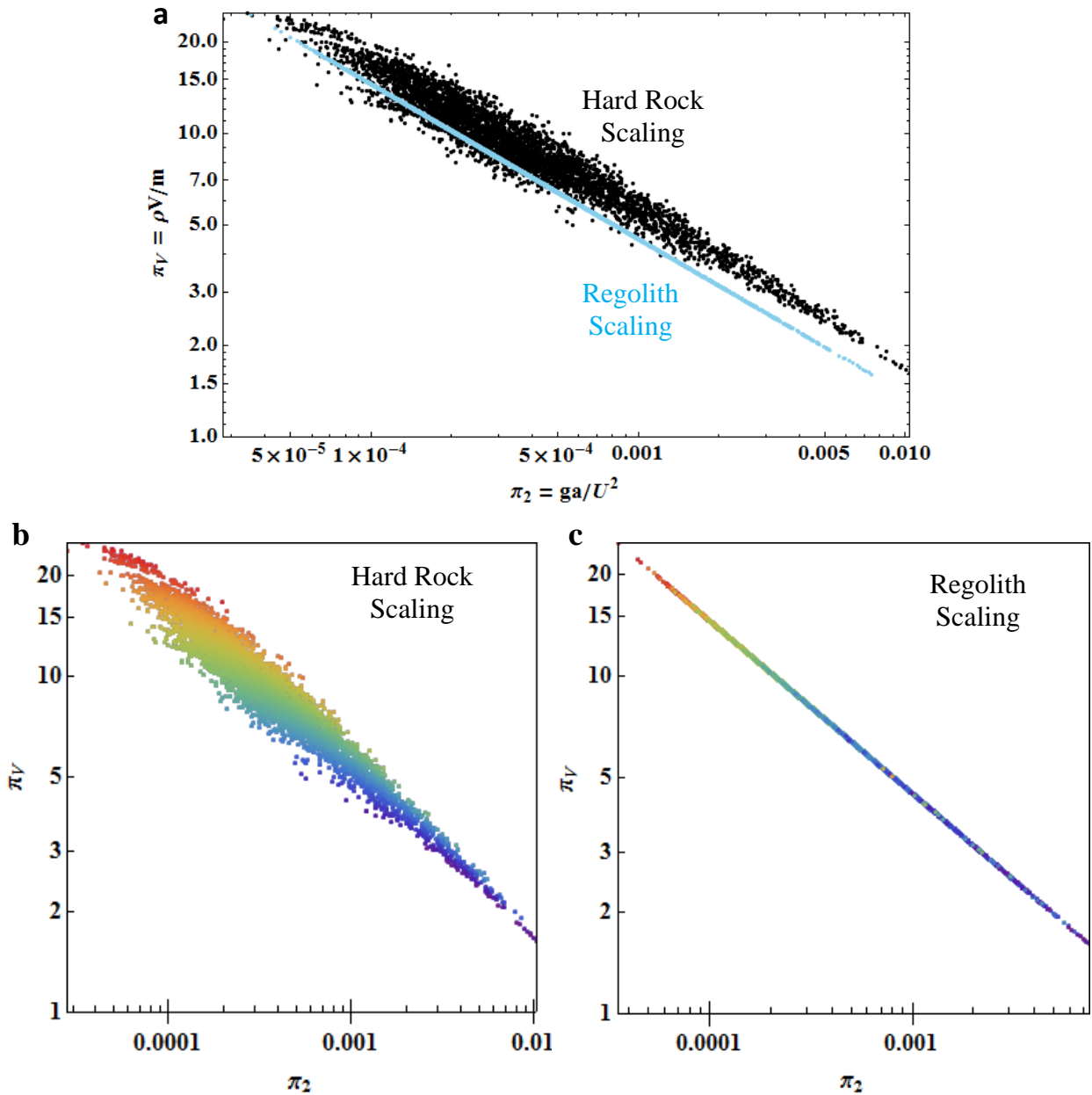

**Figure S4. Example of $\pi_2$ and $\pi_V$ values for Copernicus. a,** The gravity scaled sizes ($\pi_2$) and cratering efficiencies ($\pi_V$) for scaling the Copernicus secondary craters with "hard rock" material parameters (black circles) and "regolith" material parameters (blue squares). The dispersion for the "hard rock" material parameter scaling shows that some fragments are scaled in the transition between the strength and the gravity regimes. The points for "regolith" parameters fall on a line representing the gravity regime for this material, lunar gravity, and the velocity of the secondary fragments for each secondary crater. See Fig. S5 for all secondary fields. **b,** Copernicus data points scaled with hard rock parameters, color-coded by velocity (same data as black points in a). The velocities colors range linearly from ~365 m s$^{-1}$ in purple (lower right) to ~1055 m s$^{-1}$ in red (upper left). Higher velocity impacts generally have higher cratering efficiencies. **c,** Data plotted in a similar manner to b, but for regolith material parameters (same data as blue points in a). The combination of fragment size and velocity determines its cratering efficiency according to the scaling equations, thus the same cratering efficiency can be produced by different combinations of fragment size and velocity. Many points (~4565) are collapsed onto one line in this Figure, thus only a portion of the points are visible.

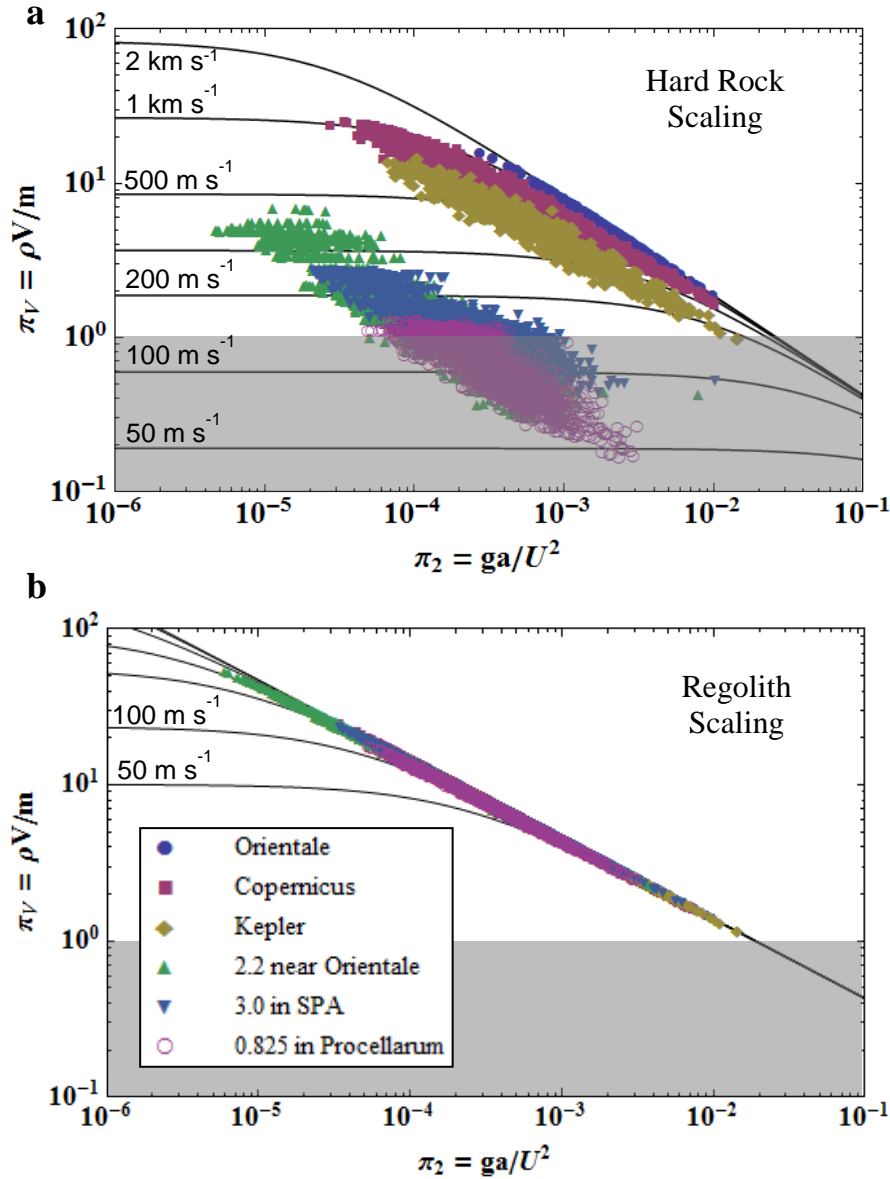

**Figure S5. Scaling for hard rock and regolith – comparison between secondary fields.** The $\pi_2$ and $\pi_V$ values for each secondary field are shown. Two endmember materials of hard-rock and regolith are shown, and the actual material properties of the lunar surface likely fall somewhere between. We selected what we deem to be the most applicable endmember for scaling as discussed in the main text: hard rock for the larger secondary craters around Orientale, Copernicus, and Kepler, and regolith for the smaller secondaries around the unnamed 3.0, 2.2, and 0.8 km primaries. All six secondary fields are represented on both **a** and **b** for comparison. Similar to Fig. S3, the lower portion of these figures ($\pi_V <$

1) is greyed out because secondary craters with the observed morphologies are unlikely to have formed at the low velocities indicated in hard rock.

## S3. Comparison to previous lunar studies

*Vickery* (1986; 1987) conducted a study of secondary craters focusing on ejecta fragment size-velocity distributions with Lunar Orbiter data for 5 secondary crater fields. Secondary crater measurements themselves were not presented, but derived fragments and SVDs were estimated. Her estimates for SVD parameters are compared to the present work in Fig. S6. Hirata and Nakamura (2006) measured secondary fields around Tycho (22 km in diameter), and Hirase et al. (2004) measured around Kepler (31 km) and Aristarchus (40 km). Neither study fit the upper envelope of the SVDs presented, but did compare their work favorably to Vickery's fits.

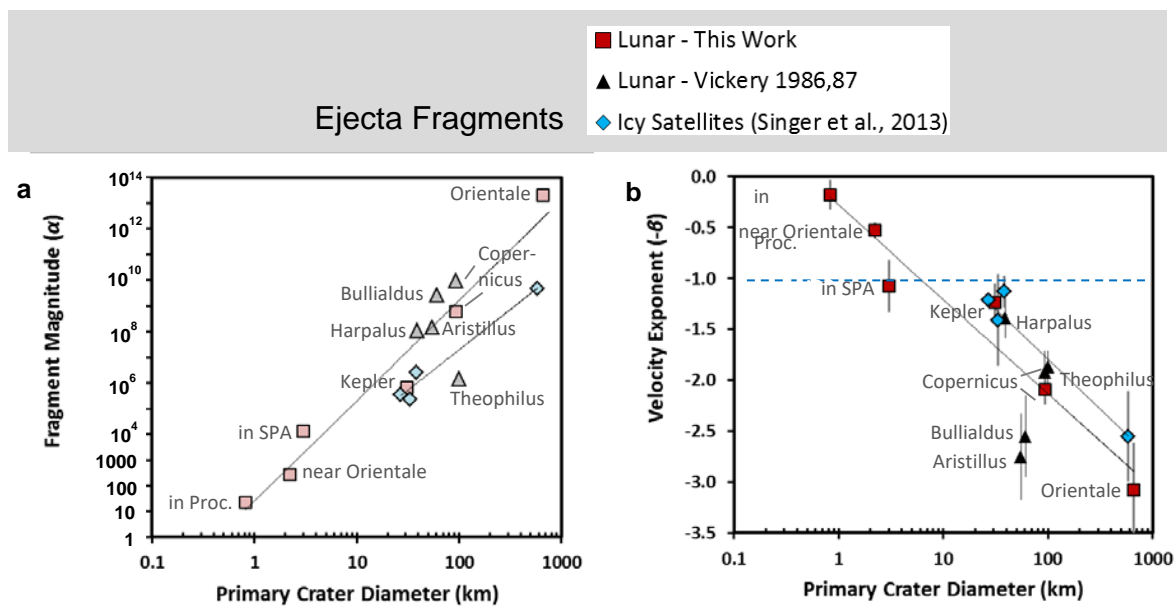

**Figure S6**. **Ejecta fragment SVD parameters** compared to the only previous lunar work on this topic by Vickery (1986); Vickery (1987). The data is generally well aligned with the present work, with a few outliers that will be investigated in future work. There is no obvious reason why Theophilus should have small ejecta fragments for its size, or why Aristillus and Bullialdus should have such steep velocity exponents. None of these impacts appear to be more oblique than the average 45° (no strong asymmetries to the ray or ejecta patterns). Theophilus secondaries are somewhat older/more degraded and occur mostly in highlands material, which may make them more difficult to map.

*O'Keefe and Ahrens* (1985; 1987) conducted a semi-empirical study where they derived equations for fragment mass and diameter as a function of total ejected mass, and the velocity dependence of the fragment distributions. Explosion cratering and empirical data were utilized to estimate a value for the power-law velocity dependence. Their analysis resulted in an estimate of a 30 m fragment being ejected from a 50 km impact on the Moon, similar to that predicted by our work. They also warned, however, that the power-law exponent was not well constrained, and changing the value could easily result in an order-of-magnitude difference in the results.

The following studies considered various aspects of lunar or terrestrial secondary cratering and ejecta, but did not characterize size-velocity distributions. Allen (1979) documented the largest 5 secondaries around 19 primary craters and found an approximatelylinear relationship in the increasing of secondary crater size with increasing primary size ($d_{sec,max} = (0.04 \pm 0.006)\ D$. Data from our current study closely aligns with Allen's, both in terms of the max secondary size trend and the distance of the largest secondaries from the primary center. Fielder (1961) measured the size and distance from primary for 51 Copernicus and 8 Tycho secondaries. He estimated the velocities of the fragments, and from there discussed the implications for determining characteristics of surface material, and formation of v-shaped features associated with secondary craters. Guest and Murray (1971) mapped "V-features" in the ejecta facies of Copernicus and determined their opening-angles and bisector azimuths in order to constrain formation theories. They noted the presence of secondary craters at the v-feature apexes but did not measure the craters themselves. Wilhelms et al. (1978) mapped secondary craters around Orientale and Imbrium. They discussed the abundance of secondaries (encompassing craters as large as 4.4–20 km in diameter) and noted the secondary crater populations had relatively steep size-frequency distributions. Marcus (1966) and Walker (1967) also dealt with lunar size-frequency distributions and factoring in the effects of secondary craters.

Roberts (1964) summarized information about explosion crater secondaries, particularly at the Sedan test site where the buried 100 Kt thermonuclear explosion produced numerous secondaries, nine of which were excavated. These craters were produced by relatively low velocity ejecta fragments and mostly formed by compression, as the original alluvium ground surface was noted below the secondary crater in the excavations. The ejecta fragment projectiles were often found in the secondary craters, or sheared and deposited just up range. Roberts' morphological descriptions can be helpfully compared to those of secondaries on other planets. The ejecta projectiles consisted of boulders, cobbles, or rock/soil fragments, man-made structural pieces, compacted and comminuted material from near the impact site, or discrete but unconsolidated masses of crater ejecta. These would be consistent with the idea that larger secondary craters on the Moon could be made up of discrete but fractured/fragmented masses of the crustal material.

## S4. Confidence intervals for 99[th] quantile fits

We calculate 95% confidence intervals as a measure of uncertainty in the 99[th] quantile fits to the data. The confidence intervals for the estimated fragment sizes (for the hard rock material parameters) are shown in Fig S7. These are also generated by bootstrapping, but in this case, we randomly sample the residuals (the distance between the data points and the 99[th] quantile regression power law on the y-axis) with replacement 1000 times. The randomly sampled residuals are added to the y-values (fragment size as shown in Fig. S7) on the quantile regression line that correspond to each x-value in the data (velocities for each data point as shown in Fig. S7) to generate 1000 bootstrapped datasets. We then estimate the power-law parameters ($\alpha$ and -$\beta$) with quantile regression for each of the 1000 bootstrapped datasets and take the 0.025 and 0.975 quantile of the parameters (the equivalent of the 25th position and the 975th positions from an ordered list of each parameter calculated for the 1000 datasets). The 0.025 and 0.975 quantiles correspond to the lower and upper bound of the 95% confidence interval, respectively.

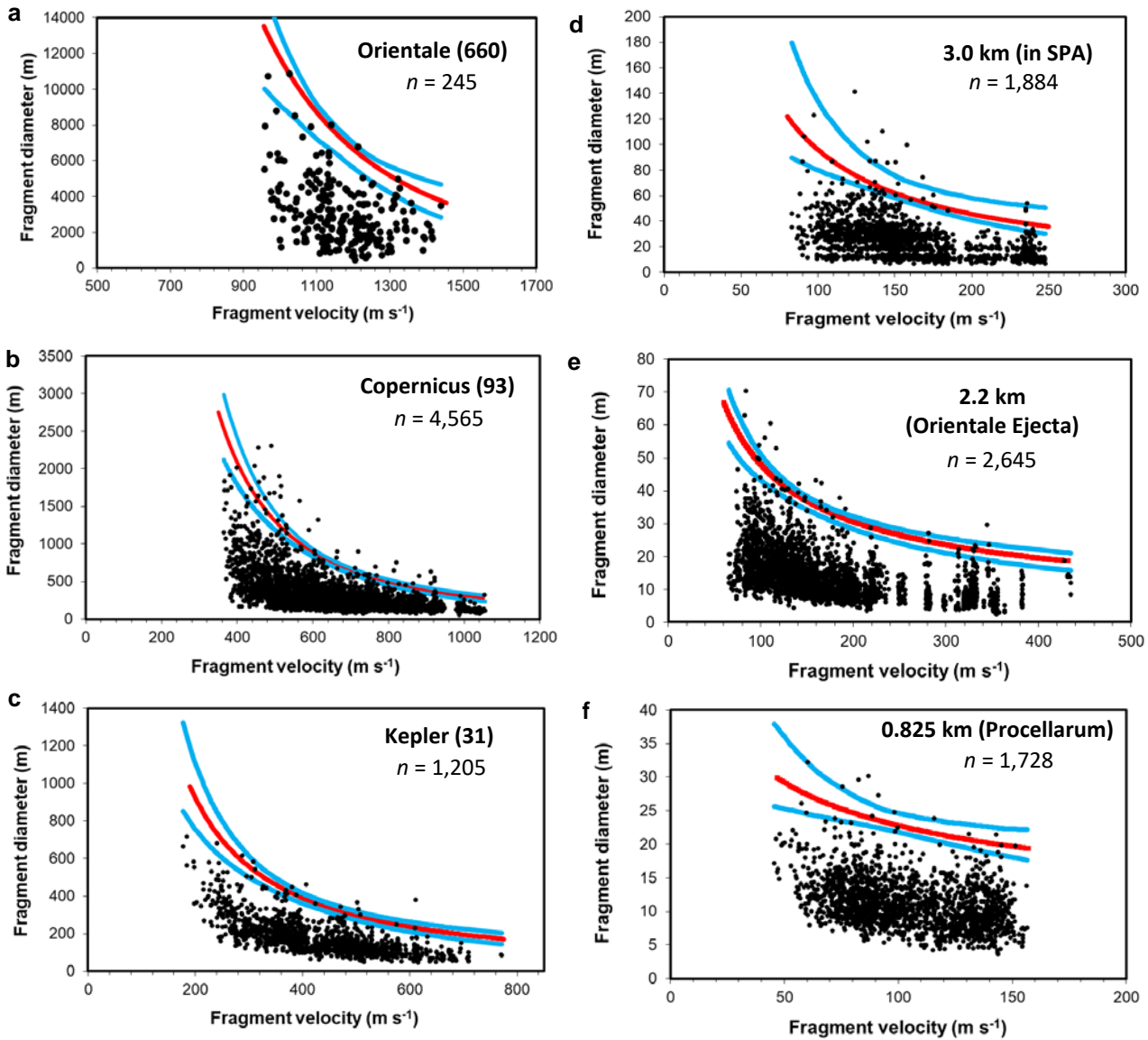

**Figure S7. Confidence intervals for ejecta fragment SVDs.** The upper and lower blue lines represent the confidence interval for the 99th quantile fits (central red lines, same as those shown in Fig. 11) for the hard rock material parameters. See accompanying text in section S4 for the description of how these confidence intervals were generated.

## Caption for Table S2

**Table S2.** Is available at https://doi.org/10.6084/m9.figshare.11299319. Table S2 is an excel spreadsheet with six tabs of data for the six secondary crater fields presented in this paper. It includes the diameters (in km) and locations (center latitude and longitude in degrees) for each secondary crater presented in the paper.